\documentclass[final,5p,times,twocolumn]{elsarticle}
\journal{Combustion and Flame}

\usepackage{graphicx}
\usepackage{grffile}
\usepackage{stfloats}
\usepackage{color}
\usepackage{rotating}
\usepackage{bm}

\usepackage{amsmath}
\usepackage{amssymb}
\usepackage{amsbsy}
\usepackage[mathscr]{euscript}
\usepackage{enumitem}

\usepackage{natbib}
\biboptions{sort&compress}

\begin{document}

\begin{frontmatter}
\title{Detonation initiation by compressible turbulence thermodynamic fluctuations}

\author[fir]{Colin A.~Z.~Towery\corref{cor1}}
\ead{colin.towery@colorado.edu}

\author[thir]{Alexei Y.~Poludnenko}
\author[fir]{Peter E.~Hamlington}

\address[fir]{University of Colorado Boulder, Department of Mechanical Engineering, UCB 427, 1111 Engineering Drive, Boulder, CO 80309, USA}
\address[thir]{Texas A\&M University, Department of Aerospace Engineering, College Station, TX 77843, USA}
\cortext[cor1]{Corresponding author}

\begin{abstract}
Theory and computations have established that thermodynamic gradients created by hot spots in reactive gas mixtures can lead to spontaneous detonation initiation. However, the current laminar theory of the temperature-gradient mechanism for detonation initiation is restricted to idealized physical configurations. Thus, it only predicts conditions for the onset of detonations in quiescent gases, where an isolated hot spot is formed on a timescale shorter than the chemical and acoustic timescales of the gas. In this work, we extend the laminar temperature-gradient mechanism into a statistical model for predicting the detonability of an autoignitive gas experiencing compressible isotropic turbulence fluctuations. Compressible turbulence forms non-monotonic temperature fields with tightly-spaced local minima and maxima that evolve over a range of timescales, including those much larger than chemical and acoustic timescales. We examine the utility of the adapted statistical model through direct numerical simulations of compressible isotropic turbulence in premixed hydrogen-air reactants for a range of conditions. We find strong, but not conclusive, evidence that the model can predict the degree of detonability in an autoignitive gas due to turbulence-induced thermodynamic gradients.
\end{abstract}


\end{frontmatter}

\section{Introduction\label{sec:intro}}

Turbulence compressibility, autoignitive combustion, nonlinear interactions between turbulence and chemistry, and transitions to detonation are all fundamental aspects of many different high-speed reacting flows, from human-scale engineered flows such as supersonic combustion ramjet engines (scramjets) \cite{Urzay2018} to astrophysical flows such as Type 1a supernovae \cite{Oran2007}. It is now well-established that thermodynamic gradients within localized \emph{hot spots} (also termed \emph{exothermic centers}) can lead to the direct initiation of detonations in auto-igniting gas mixtures \cite{Zeldovich1970, Lee1978, Zeldovich1980, Bradley1996}. It is also established that nonlinear turbulence-chemistry interactions alone can induce deflagration-to-detonation transition (DDT) in high-speed turbulent flames. This can occur via the same localized mechanism of hot-spot autoignition \cite{Khokhlov1991, Khokhlov1997}, where the necessary thermodynamic gradient is created by the turbulent flame rather than an external source. Alternatively, DDT can be induced through a self-reinforcing amplification of an initial large-scale pressure wave generated during a self-acceleration event of an unsteady turbulent flame propagating with a speed above that of a Chapman-Jouguet (CJ) deflagration \cite{Poludnenko2011a, Poludnenko2015}. All three mechanisms --- autoignitive hot-spot initiation, local hot-spot DDT, and turbulence-driven DDT in super-CJ flames --- rely, at least in part, on a thermomechanical feedback loop between chemical heat release and pressure in order to amplify an acoustic wave into a detonation wave.  This unsteady feedback loop is described in detail for the particular case of sub-detonable shock waves, termed \emph{shock wave amplification by coherent energy release}, or SWACER, in \cite{Lee1978}. Qualitatively, the critical difference between the detonation mechanisms is the source that initiates and drives the feedback mechanism: autoignition or turbulent deflagration.

In autoignitive gaseous flows, spatial gradients in temperature, pressure, and species mixture fractions produce spatial gradients in the chemical ignition delay time, denoted $t_\mathrm{ign}$. As long as advective and diffusive processes operating on the temperature and species gradients are slow compared to the range of ignition delay times present \cite{Khokhlov1991}, a monotonic gradient will form a \emph{spontaneous} reaction front that propagates with a speed equal to the inverse of the local ignition delay time gradient magnitude, namely $u_\mathrm{sp} = |\nabla t_\mathrm{ign}|^{-1}$. 

Neglecting any thermomechanical response of the gas to initial conditions or subsequent heat release, an isolated hot spot with a constant $u_\mathrm{sp}$ as the initial condition will form either a subsonic or supersonic spontaneous ignition wave that will transition to a conventional deflagration or detonation wave within specific regimes of the spontaneous wave-speed. These four propagation regimes are delineated by the magnitude of $u_\mathrm{sp}$ with respect to the local Chapman-Jouguet detonation speed, $D_\mathrm{CJ}$, the upstream speed of sound, $a$, and the laminar deflagration speed, $S_\mathrm{L}$ \cite{Zeldovich1980}. Only in the case where $a < u_\mathrm{sp} < D_\mathrm{CJ}$ will the spontaneous ignition wave formed from initial conditions transition to a detonation wave.

When the thermomechanical response of the gas to both the initial conditions and heat release are taken into account, a far larger range of $u_\mathrm{sp}$ can transition to a detonation wave than would be expected from an analysis of the initial conditions alone \cite{Lee1978, Khokhlov1991, Kassoy2010}. Most notably, when $S_\mathrm{L} < u_\mathrm{sp} < a$, a spontaneous subsonic ignition wave would be predicted to form and propagate through the hot spot without transitioning to another regime \cite{Zeldovich1980}. However, the thermomechanical response of the surrounding gas to the initial creation of the hot spot will form an acoustic wave, independent of the ignition wave \cite{Lee1978, Kassoy2010}. Due to the monotonically-decreasing speed of sound, the downstream pressure increase due to heat release from the subsonic ignition wave will propagate upstream faster than the leading acoustic wave, allowing the trailing subsonic ignition wave to amplify the leading acoustic wave. In turn, the ignition-delay-time gradient between the acoustic and ignition waves will be smaller than in the initial conditions, and therefore the ignition wave will accelerate towards the acoustic wave. Consequently, the heat release rate will increase within the reaction zone of the wave. This process is self-reinforcing, in the same manner as SWACER feedback mechanism \cite{Lee1978}, and will eventually accelerate the ignition wave to supersonic speeds. At this point, the ignition wave will overtake and coalesce with the leading acoustic wave and transition to a detonation wave. This broader temperature-gradient-based mechanism for the direct initiation of detonation waves has been extensively validated from both an analytical perspective, through asymptotic analysis and reduced order models, and from a numerical perspective, through one- and two-dimensional (1D and 2D, respectively) numerical simulations of simplified hot spot configurations. The reader is directed to \cite{Bartenev2000, Kapila2002} for comprehensive reviews of the foundational literature.

Both Khokhlov \cite{Khokhlov1991} and Bradley \cite{Bradley1996} numerically studied the thermomechanical response of 1D hot spots to chemical heat release in order to discern the critical range of hot-spot radii and temperature-gradient magnitudes that lead to detonation initiation. In addition to the ratio $a/u_\mathrm{sp}$, these studies determined that detonation initiation also depends on the ratio of the characteristic time scale of exothermic heat release, $t_\mathrm{exo}$, within the hot spot to the acoustic residence time of the hot spot, ${r_0}/a$, where $r_0$ is the initial radius prior to thermomechanical relaxation. The importance of this \emph{acoustic-exothermicity coupling} was expanded upon in subsequent studies \cite[e.g.,][]{Gu2003, Radulescu2013, Regele2013, Kurtz2014, Bates2016}, establishing that the detonability of an isolated hot spot can be characterized by two non-dimensional parameters, given as
\begin{align}
    \xi &{}= \frac{a}{u_\mathrm{sp}} = a|\nabla t_\mathrm{ign}| \approx
             a t_\mathrm{ign}\frac{T_\mathrm{a}}{T^2}|\nabla T|_\mathrm{hs}\,, \label{eq:xi}\\
    \zeta &{}= \frac{r_0}{a t_\mathrm{exo}}\,, \label{eq:zeta}
\end{align}
where $T$ is the temperature, $T_\mathrm{a} = {\mathrm{d}(\ln t_\mathrm{ign})}/{\mathrm{d}(1/T)}$ is the effective activation temperature of the chemistry, $|\nabla T|_\mathrm{hs}$ is the average or constant temperature gradient of the hot spot, and $t_\mathrm{ign}$, $t_\mathrm{exo}$, $T$, and $T_\mathrm{a}$ are each evaluated from the initial conditions at the radial midpoint, $r_0/2$. It should be noted that no assumptions have been made about the explosive or autoignition limits of the reactant mixture in deriving $\xi$ and $\zeta$. The only assumptions are that the two chemical timescales, $t_\mathrm{ign}$ and $t_\mathrm{exo}$, are quantifiable, or alternatively that $|\nabla t_\mathrm{ign}|$ is well approximated by $t_\mathrm{ign}T_\mathrm{a}|\nabla T|/T^2$.

As described by Gu \emph{et al.}~\cite{Gu2003}, the acoustic-exothermicity coupling, $\zeta$, predicts the detonability of the hot spot according to the following criteria:
\begin{itemize}[itemsep=0pt,leftmargin=*]
    \item If $\zeta \gg 1$, then the spontaneous ignition wave will form much faster than an acoustic wave can exit the hot spot, and the two waves will coalesce and transition into a fully-developed detonation wave before both waves exit the region of negative speed-of-sound gradient in the hot spot that enables the self-reinforcing thermomechanical coupling between the waves.
    \item If $\zeta \approx 1$, then the ignition wave will form in the same time that the acoustic wave will have crossed the initial hot spot radius, and whether a detonation wave forms will be strongly dependent on the value of $\xi$, because the continuous thermomechanical relaxation of the hot spot \emph{may} continue to decelerate the acoustic wave long enough to allow for the self-reinforcing thermomechanical coupling to bring the two waves together.
    \item If $\zeta \ll 1$, then the ignition wave will not form before the acoustic wave has left the hot spot and ceased to decelerate, and none of the heat release within the ignition wave will be capable of amplifying the acoustic wave in a self-reinforcing manner.
\end{itemize}
In accordance with these criteria, the \emph{acoustic-induction coupling}, $\xi$, has upper and lower detonability limits that correspond to $a/S_\mathrm{L}$ and $a/D_\mathrm{CJ}$ in the asymptotic limit $\zeta \gg 1$, whereas at $\zeta\approx1$, the upper and lower detonability limits contract to a single point, typically in the range $2 < \xi < 7$, forming a ``detonation peninsula'' within the $\zeta-\xi$ parameter space \cite{Bates2016}. 

The temperature-gradient mechanism of detonation initiation is, however, limited to the prediction of detonations in localized and isolated hot spots that are formed on timescales shorter than, or comparable to, chemical and acoustic timescales. In the case of highly turbulent autoignitive flows, by contrast, turbulence compressibility can generate non-monotonic temperature fields with tightly-spaced minima and maxima that vary over a wide range of length and time scales, including those much larger than chemical and acoustic length and time scales. As such, there is currently no \emph{predictive} model for the initiation of detonations by compressible turbulence fluctuations in an autoignitive flow. Therefore, in this work, we adapt these governing non-dimensional parameters of the laminar temperature-gradient mechanism into a statistical model for the \emph{a priori} prediction of spontaneous detonation initiation by thermodynamic gradients formed from flow-field fluctuations in compressible homogeneous isotropic turbulence (HIT). We then use three-dimensional (3D) direct numerical simulations (DNS) of compressible HIT in premixed hydrogen-air reactants over a small --- but critical --- range of conditions to examine the utility of the adapted statistical model for predicting compressible turbulence detonability.

The present focus on autoignitive HIT is motivated by prior studies of velocity and thermodynamic fluctuations in both non-reacting and reacting turbulent flows. For non-reacting equilibrium turbulent flows at very high Reynolds numbers, it is widely hypothesized that the smallest scales behave in a universal manner independent of any particular large-scale flow geometry, and, as a consequence, the turbulent fluctuations must be statistically homogeneous and isotropic \cite{Frisch1995, Sagaut2008, Schumacher2014}. In this context, the interactions of HIT with chemistry can be viewed as a model problem that approximates, on a statistical basis, the dynamics of high-pass-filtered turbulent fluctuations in a material volume of reacting fluid advecting with the bulk flow of any sufficiently high Reynolds number combustion process that has the same initial reactant and final product states \cite{Poludnenko2010, Hamlington2011a, Aspden2011, Hawkes2012}. Similarly, the effect of compressible HIT thermodynamic fluctuations on autoignition can be isolated from the effect of turbulent fluctuations in reactant mixture fractions by analyzing fully premixed reactants. In this case, the assumptions of small-scale homogeneity and premixed reactants allow for highly detailed analyses of turbulence-chemistry interactions, including multi-scale and cross-scale interactions \cite{Kolla2014, Poludnenko2015, Towery2016, Obrien2017, Kim2018, Whitman2019}.

The present work draws upon previous 2D DNS studies of superimposed temperature and turbulent kinetic energy fluctuations at conditions relevant to homogeneous charge compression ignition engines to study the importance of turbulent mixing and diffusive timescales on the development of reaction fronts \cite{Sankaran2005, Chen2006, Hawkes2006, Yoo2011, Im2015}, and the dynamical impact of chain-branching-regulated chemical kinetics, including negative-temperature-coefficient behavior, compared to the thermally-regulated one-step Arrhenius chemistry models \cite[e.g.,][]{Liberman2012, Dai2015}. Additionally, Yu and Bai \cite{Yu2013} compared saddle-shaped and spherically-curved reaction fronts, as well as the effects of 3D turbulence on autoignition, in what may be the only computational study to date that utilizes 3D DNS with detailed chemistry to focus exclusively on premixed autoignition waves. Furthermore, the present work draws upon previous studies of 3D compressible HIT, particularly 3D DNS studies of compressible turbulence thermodynamics \cite{Wang2013e, Donzis2013, Jagannathan2016}. However, no previous study has performed DNS of turbulent premixed autoignition where the controlling temperature gradients were a direct result of compressible turbulence fluctuations. This intrinsic, turbulence-induced mechanism of detonation initiation is the primary focus of the present study. 

This paper is organized as follows. First, we extend the dimensional analysis for detonation initiation in isolated hot spots to turbulence-induced thermodynamic fluctuations in compressible mixtures. Next, we describe details of the numerical simulation algorithms and procedures. This is followed by a detailed analysis of the results, with conclusions presented at the end.

\section{A Predictive Statistical Model for Turbulence Detonability\label{sec:model}}
In a turbulent autoignitive flow, turbulence compressibility generates spatial and temporal variations in temperature that evolve over timescales both smaller and potentially much larger than chemical and acoustic timescales. This would seem to violate much of the theoretical basis for the temperature-gradient mechanism of detonation initiation in quiescent gases with isolated hot spots, outlined in the previous section. However, due to the statistically-predictable structure and scale of turbulence fluctuations, we can characterize and predict, \emph{a priori}, the global system detonability of the auto-igniting gas using suitable spatial averages of the thermodynamic fluctuations as arguments in a dimensional analysis of the system detonability. This is carried out by replacing the laminar hot-spot temperature, temperature gradient, and radius in Eqs.~\eqref{eq:xi} and \eqref{eq:zeta} with the mass-average temperature, mass-weighted root-mean-square (rms) temperature gradient, and a temperature-field equivalent to the Taylor length scale, respectively. These substitutions yield new turbulence-based expressions for $\xi$ and $\zeta$, denoted $\xi_\mathrm{t}$ and $\zeta_\mathrm{t}$, given as
\begin{align}
    \xi_\mathrm{t}   &{} = a_\mathrm{rms}\tau_\mathrm{ign}\frac{T_\mathrm{a}}{\{T\}^2}|\nabla T|_\mathrm{rms}\,,\label{eq:xt} \\
    \zeta_\mathrm{t} &{} = \frac{2\lambda_T}{a_\mathrm{rms}\tau_\mathrm{exo}}
                     = \frac{2T'}{a_\mathrm{rms}\tau_\mathrm{exo}|\nabla T|_\mathrm{rms}}\,, \label{eq:zt}
\end{align}
where $\{\cdot\} = \langle \rho \cdot \rangle / \langle \rho \rangle$ denotes a mass-weighted spatial average, $\langle\cdot\rangle$ denotes a volume-weighted spatial average, $a_\mathrm{rms}^2 = \{\gamma P/\rho\}$ is the rms speed of sound, and $\lambda_T = T'/|\nabla T|_\mathrm{rms}$ is the temperature-field Taylor scale. The standard deviation of the temperature fluctuations, $T'$, and the rms temperature gradient, $|\nabla T|_\mathrm{rms}$, are 
\begin{align}
     T' & = \left\{\left(T - \{T\}\right)^2\right\}^{1/2}\,, \\
     |\nabla T|_\mathrm{rms} & = \left\{\frac{\partial T}{\partial x_j}\frac{\partial T}{\partial x_j}\right\}^{1/2}\,.
\end{align}

In deriving $\zeta_\mathrm{t}$, we assumed that a linearized approximation of an effective ``rms hot spot'' has a radius $r_0 = 2\lambda_T$ and varies from a maximum temperature of $\{T\}+T'$ at its center down to $\{T\}-T'$ in the surrounding cold gas, giving a piecewise-constant temperature gradient magnitude of $|\nabla T|_\mathrm{rms}$. Similarly, $\xi_\mathrm{t}$ represents an effective ``rms spontaneous velocity'' for the entire simulation domain. Each thermodynamic moment (i.e., mean, and rms or standard deviation) is assumed to be computed for the turbulent flow field at the initiation of reactions in the simulation (see \S\ref{sec:methods}, below), while the characteristic ignition delay time, $\tau_\mathrm{ign}$, exothermicity time, $\tau_\mathrm{exo}$, and effective activation temperature, $T_\mathrm{a}$, are computed for a homogeneous constant volume reaction with initial conditions $\langle P\rangle$ and $\{T\}$.

Although it is possible to compute rms values for $\zeta$ and $\xi$ directly from a simulation, \emph{a posteriori}, or even obtain complete probability distribution functions as a function of time within the unreacted regions of the domain, such a diagnostic approach to assessing turbulence detonability cannot provide the basis for an \emph{a priori} predictive model. By formulating the single characteristic values $\xi_\mathrm{t}$ and $\zeta_\mathrm{t}$ as functions solely of the initial thermodynamic conditions, dimensional analysis of inert compressible HIT can be used to compute \emph{a priori} estimates of $T'$ and $|\nabla T|_\mathrm{rms}$ for any given simulation conditions, as outlined in Section \ref{ssec:nondim}. 

It is also possible to use only volume-weighted statistics to derive $\xi_\mathrm{t}$ and $\zeta_\mathrm{t}$, which could simplify comparison to experimental investigations. In fact, at low $Ma_\mathrm{t}$, the mass-weighted and volume-weighted averages deviate only slightly. However, the disagreement between the two averages increases significantly with $Ma_\mathrm{t}$, as reported in \cite{Jagannathan2016}. More importantly, the use of a mass-average for temperature statistics is the natural result of solving a conservative formulation of the reacting flow equations for an ideal gas, given in \S\ref{ssec:nse}. In particular, if one knows the volume-averaged pressure and density in either a computational simulation or experimental apparatus, and the mixture-averaged molecular weight is uniform in space (i.e., the gases are fully premixed but unreacted), then $\{T\} = \langle P\rangle/(\langle\rho\rangle R)$ is easily obtained. Due to this algebraic consistency in both the governing differential equations and the ideal gas law, the mass-average temperature is much preferred to the volume-average temperature as an input to a computational model for compressible turbulence where $Ma_\mathrm{t}$ is a free parameter.

We hypothesize that the degree of detonability of temperature fluctuations induced by compressible turbulence can be reliably predicted by comparing $\zeta_\mathrm{t}$ to unity within some empirical upper and lower bounds of $\xi_\mathrm{t}$, as with laminar hot spots (see the criteria outlined in the previous section). In the following sections, we test this hypothesis using five new DNS cases of auto-igniting stoichiometric hydrogen-air mixtures subject to compressible isotropic turbulence in which $\zeta_\mathrm{t}$ is varied while $\xi_\mathrm{t}$ is held nearly constant at an intermediate value likely to favor detonability.

\section{Methods\label{sec:methods}}

\subsection{Physical model of turbulent autoignition\label{ssec:nse}}
In the present simulations, premixed turbulent combustion is represented using a system of partial differential equations derived from the compressible Navier-Stokes equations (which describe conservation of mass, momentum and total energy) subject to initial and boundary conditions, as given by
\begin{align}
    \frac{\partial\rho}{\partial t} =       &{} - \frac{\partial}{\partial x_j}(\rho u_j)\,, \label{eq:mass}\\
    \frac{\partial(\rho u_i)}{\partial t} = &{} - \frac{\partial}{\partial x_j}(\rho u_iu_j  + P\delta_{ij} - \sigma_{ij})
                                    + F_i\,, \label{eq:momentum}\\
    \frac{\partial(\rho E)}{\partial t} =   &{} - \frac{\partial}{\partial x_j}\bigg(\rho E u_j  + Pu_j - q_j - u_i\sigma_{ij} \bigg) \nonumber\\
                                &{} + \rho \Delta h_{\mathrm{f},k}\dot{\Omega}_k  + u_j F_j\,, \label{eq:energy} \\
    \frac{\partial(\rho Y_k)}{\partial t} = &{} - \frac{\partial}{\partial x_j}\bigg(\rho Y_ku_j - \mathcal{D}_{kj} \bigg)
                                    + \rho\dot{\Omega}_k\,, \label{eq:species}
\end{align}
where twice-repeated indices in a term imply summation, $\rho$ is the density, $u_i$ is the velocity vector, $Y_k$ is the mass fraction of the $k$th chemical species, $E$ is the specific total energy, $P$ is the pressure, $T$ is the temperature, $\Delta h_{\mathrm{f},k}$ is the specific enthalpy of formation of the $k$th species, and $\dot{\Omega}_k$ is the production rate of the $k$th species. The viscous stress tensor, $\sigma_{ij}$, diffusive heat flux, $q_j$, and diffusive species flux, $\mathcal{D}_{kj}$, are
\begin{align}
    \sigma_{ij} & = 2\mu(S_{ij}-\frac{1}{3}\theta\delta_{ij}) + \mu_b\theta\delta_{ij}\,, \\
    q_j & = \kappa\frac{\partial T}{\partial x_j} - \rho h_k\mathcal{D}_{kj}\,, \\
    \mathcal{D}_{kj} & = \rho D_k\frac{Y_k}{X_k}\frac{\partial X_k}{\partial x_j}\quad\text{(no summation over }k\text{)}\,, \label{eq:chemflux}
\end{align}
respectively, where the multiply-repeated index $k$ does not imply summation in Eq.~\eqref{eq:chemflux}. Here, $S_{ij}= \tfrac{1}{2}(\partial u_i/\partial x_j + \partial u_j/\partial x_i)$ is the strain rate tensor, $\theta = S_{ii}$ is the dilatation, $\mu$, $\mu_b$, and $\kappa$ are the mixture-averaged shear viscosity, bulk viscosity, and thermal conductivity, respectively, and $D_k$, $h_k$, and $X_k$ are the mixture-averaged mass diffusivity, specific enthalpy, and mole fraction for the $k$th species, respectively. The fluid is modeled as a thermally-perfect ideal gas, where thermodynamic variables are related to each other by the ideal gas equation of state, $P = \rho R_\mathrm{u}T/W_\mathrm{mix}$, where $R_\mathrm{u}$ is the universal gas constant. Additionally, a time-varying body force, $F_i$, is applied in order to generate and maintain isotropic turbulence fluctuations within the periodic domain; further details of this forcing are outlined in the next section.

The physical domain of the HIT simulations examined here is triply-periodic and is therefore a closed, adiabatic, and isochoric control volume. As a result, without turbulence fluctuations in the thermodynamic state of the premixed autoignitive ideal gas mixture, the entire system would undergo a homogeneous constant-volume (CV) explosion. With the turbulence fluctuations, the spatially-averaged system remains isochoric, but with a different temporal evolution than a homogeneous CV reaction. 

Locally and instantaneously, each infinitesimal parcel of reacting gas may undergo any kind of change in pressure and specific volume during combustion, including constant-pressure (isobaric) auto-ignition and detonation cycles that do not correspond to ideal Chapman-Jouguet (CJ) detonation speeds and Zel'dovich-von Neumann-D{\"o}ring (ZND) wave structure (see \cite{Kuo1986, Fickett2000, Kao2008}, e.g., for descriptions of CJ and ZND theory). Despite this, all combustion modes (i.e., autoignition, deflagration, and detonation) can be universally characterized by the evolution of the \emph{total thermicity}, $\dot{\omega}$, which has units of s$^{-1}$ and expresses the total thermodynamic effect of chemical reactions, including both the volume increase at constant pressure due to the change in the number of molecules, and the heat release due to the breaking of chemical bonds \cite{Fickett2000, Kao2008}. For an ideal gas, $\dot{\omega}$ is given as
\begin{equation}
    \dot{\omega} = \left(\frac{W_\mathrm{mix}}{W_k} - \frac{h_k}{c_P T}\right)\dot{\Omega}_k\,,
\end{equation}
where, as before, the twice-repeated index $k$ implies summation, $c_P$ is the mixture-averaged specific heat capacity at constant pressure, $W_\mathrm{mix}$ is the mixture-averaged molecular weight, and $W_k$ is the molecular weight of the $k$th species in the gas mixture. 

Within this context, we can define the ignition delay time, $t_\mathrm{ign}$, to be the time required to reach the maximum total thermicity from the upstream or initial condition, and the exothermicity time, $t_\mathrm{exo} = t_\mathrm{shm}-t_\mathrm{fhm}$, as the difference between the time that the thermicity first rises to 50\% of its maximum [i.e., the first half-max (fhm) state] and the time that the thermicity subsequently drops back below 50\% of its maximum [i.e., the second half-max (shm) state] \cite{Kao2008}. The reaction-zone width of a propagating combustion wave is equivalently defined as the distance in space between these two states, i.e., $\delta_\mathrm{exo} = |\bm{x}_\mathrm{shm} - \bm{x}_\mathrm{fhm}|$.

Chemical reactions in this study are modeled using a multi-species kinetic mechanism for hydrogen combustion based on the 2014 San Diego mechanism \cite{Sanchez2014}. This mechanism includes 21 reactions involving 8 reacting species (i.e., H, H$_2$, O, O$_2$, OH, H$_2$O, HO$_2$, and H$_2$O$_2$) and the inert N$_2$. Thermodynamic properties are given by polynomial functions of temperature in the NASA seven coefficient format \cite{McBride1993}. Pure species viscosity coefficients and binary diffusion coefficients are computed from standard kinetic theory given in Hirschfelder \emph{et al.}~\cite{Hirschfelder1954}, while thermal conduction coefficients of pure species are computed from expressions given in Warnatz~\cite{Warnatz1982}. Mixture-averaged conduction, shear viscosity, and bulk viscosity coefficients are computed from averaging formulas of order $1/4$, $6$, and $1/2$ respectively \cite{Ern1994}, while mixture-averaged species diffusion coefficients are computed using the same method as in the TRANSPORT library \cite{Kee1986}.

\subsection{Details of the direct numerical simulations\label{ssec:dns}}
Equations \eqref{eq:mass}-\eqref{eq:species} are solved numerically on a uniform grid by operator splitting the reaction and forcing terms from the conservative advection and diffusion terms. Advection and diffusion are advanced forward in time by the fully unsplit corner transport upwind finite-volume scheme with piecewise-parabolic monotone spatial reconstruction and the HLLC Riemann solver \cite{Colella1990, Gardiner2008} implemented in the code \texttt{Athena-RFX} \cite{Stone2008, Poludnenko2010,Poludnenko2015}.

Specialized code for the chemical reaction source terms is generated from a CHEMKIN \cite{Kee1996} input file in a preprocessing step that generates the rates of change and analytical Jacobian \cite{Perini2012} of the species mass fractions and temperature due to chemical reactions. At run time, \texttt{Athena-RFX} reads CHEMKIN and TRANSPORT input files to obtain thermodynamic and transport property data and tabulates pure species transport properties, as well as forward and reverse reaction rate constants and their temperature derivatives, for subsequent evaluation by interpolation \cite{Hamlington2017}. The stiff system of equations for the chemical kinetics is integrated using the non-iterative, single step, semi-implicit ODE integrator YASS \cite{Khokhlov2012}. It does not employ any approximations to the Jacobian matrix, it explicitly conserves species mass fractions and total energy, and it provides an excellent balance between accuracy and efficiency, which is critical for large-scale DNS.

Isotropic turbulence is sustained in each DNS case by introducing a body force, $F_i$, in the governing equations for the momentum and specific total energy [Eqs.~\eqref{eq:momentum} and \eqref{eq:energy}, respectively]. This body force is implemented as a large-scale, isotropic perturbation momentum field $\delta f_i$ that is added discretely after each time-step \cite{Petersen2010}. Moreover, the effective kinetic energy of $\delta f_i$ is modulated to provide a constant rate of energy injection, $\varepsilon_\mathrm{inj}$. In this context, $F_i = \delta f_i/\Delta t$, where $\Delta t$ is the current simulation time-step. The forcing method amplifies the large-scale motions of the momentum field, and thus $\delta f_i$ is formed deterministically from the solution momentum field $\rho u_i$ as
\begin{equation}
    \delta f_i = \frac{\langle\rho\rangle\varepsilon_\mathrm{inj}}
                      {\left\langle u_i
                        \left(\sqrt{\rho}\delta w^s_i - \langle\sqrt{\rho}\delta w^s_i\rangle\right)
                       \right\rangle}
                 \left(\sqrt{\rho}\delta w^s_i - \langle\sqrt{\rho}\delta w^s_i\rangle\right)\,,
\end{equation}
where $w_i = \sqrt{\rho}u_i$, and $\delta w^s_i$ is computed in spectral space as
\begin{equation}
    \widehat{\delta w^s_i}(\bm{k}) =
    \begin{cases}
    	\left(\delta_{ij} - k_ik_j/|\bm{k}|^2\right)\widehat{w}_j(\bm{k})\,, & \text{if}\ 2\Delta k \leq k < 4\Delta k \\
    	0\,,                                                          & \text{otherwise}\,,
    \end{cases}
\end{equation}
where $k_i$ is the wavenumber vector (also denoted $\bm{k}$), $\Delta k = 2\pi/L$ is the spectral grid resolution, $L$ is the domain width, and $\widehat{(\cdot)}$ denotes a Fourier transform.  The perturbation momentum, $\delta f_i$, is recomputed every $t_\mathrm{vp} = \max(\Delta t, \tau_\mathrm{K}/4)$, where $\tau_\mathrm{K}$ is the Kolmogorov time scale, and is appropriately normalized to ensure a constant energy injection rate with no mean momentum addition. For well-resolved DNS at weak to moderate compressibility levels, the typical simulation time-step, $\Delta t$, will be much less than $\tau_\mathrm{K}$, and therefore limiting the refresh rate to $t_\mathrm{vp} = \tau_\mathrm{K}/4$ provides significant computational cost savings by greatly reducing the number of forward and inverse parallel fast Fourier transforms that must be computed on the solution field.

\subsection{Physical setup of the simulations\label{ssec:nondim}}

Collectively, the constant rate of kinetic energy injection from turbulence forcing, $\varepsilon_\mathrm{inj}$, the domain box size, $L$, and the initial temperature, $T_\mathrm{init}$, pressure, $P_\mathrm{init}$, and species mole fractions $X_{k,\mathrm{init}}$, fully determine the physical configuration and dimensional scaling of each simulation prior to the start of reactions. Based upon the particular turbulence forcing scheme and solution procedure used, very accurate empirical estimates of the rms velocity, $u_\mathrm{rms} = \{u_iu_i\}^{1/2}$, $\{T\}$, and $\langle P\rangle$ at the start of reactions can be made from $\varepsilon_\mathrm{inj}$, $L$, $T_\mathrm{init}$, and $P_\mathrm{init}$. 

Neglecting the initial chemical mixture, $X_{k,\mathrm{init}}$, which we hold constant across all simulations, four independent non-dimensional parameters will govern the dynamics of non-reacting compressible HIT: the Taylor-scale Reynolds number, $Re_\mathrm{t}$, turbulence Mach number $Ma_\mathrm{t}$, the characteristic ratio of specific heats, $\gamma_0 = \gamma(\{T\})$, and the Prandtl number $Pr$, where 
\begin{align}
    Re_\mathrm{t} &{}= \frac{u_\mathrm{rms}\lambda_u}{\sqrt{3}\{\nu\}} \approx u_\mathrm{rms}^2\left(\frac{5}{3\nu_0\varepsilon_\mathrm{inj}}\right)^{1/2}\,, \\
    Ma_\mathrm{t} &{}= \frac{u_\mathrm{rms}}{a_\mathrm{rms}} \approx \frac{u_\mathrm{rms}\langle\rho\rangle}{\gamma_0\langle P\rangle}\,,\\
    Pr &{}= \frac{\nu_0}{\alpha_0}\,.
\end{align}
Here, $\lambda_u$ is the (velocity-based) Taylor length scale,  $\nu_0 = \mu(\{T\})/\langle\rho\rangle$ is the characteristic kinematic viscosity, and $\alpha_0 = \kappa(\{T\})/[\langle\rho\rangle c_p(\{T\})]$ is the characteristic thermal diffusivity. 

The standard deviation of the temperature fluctuations, $T'$, and the rms temperature gradient, $|\nabla T|_\mathrm{rms}$, can then be predicted from scaling laws of these four parameters, given generically as
\begin{align}
    \frac{T'}{\{T\}} &{}= A Ma_\mathrm{t}^a Re_\mathrm{t}^c \gamma_0^e Pr^g\,,\\
    \frac{|\nabla T|_\mathrm{rms}L}{\{T\}} &{}= B Ma_\mathrm{t}^b Re_\mathrm{t}^d \gamma_0^f Pr^h\,,
\end{align}
if the leading coefficients (i.e., $A$ and $B$) and power-law exponents (i.e., $a$-$h$) were known with a high degree of certainty. In fact, previous numerical and theoretical studies of inert and calorically-perfect (i.e., with constant $\gamma$ and $Pr$) compressible HIT (see \cite{Sagaut2008,Donzis2013,Wang2017c}, and citations therein) have found that $a$ should be exactly $2$. However, $A$ and $B$ can be expected to depend on the initial chemical mixture and thermally-perfect equations of state, and no previous studies have sought a scaling law for $|\nabla T|_\mathrm{rms}$. Therefore, we made coarse estimates for these scaling laws in order to determine the required simulation input parameters for a given choice of $Ma_\mathrm{t}$, $Re_\mathrm{t}$, $\zeta_\mathrm{t}$, and $\xi_\mathrm{t}$.

Since both $\gamma_0$ and $Pr$ depend only on $\{T\}$, and we expected the initial $\{T\}$ to vary only slightly between simulations (i.e. from 900-1200 K), we assumed that $e=f=g=h=0$ for our present purposes. Furthermore, previous studies \cite{Donzis2013,Wang2017c} found that ${T'}/{\{T\}}$ is essentially independent of $Re_\mathrm{t}$, and so we assumed that $c=d=0$. The remaining exponents ($a$ and $b$) and leading coefficients ($A$ and $B$) were estimated from a series of inert HIT simulations at $Ma_\mathrm{t} = [0.1, 0.2, 0.4]$ and $Re_\mathrm{t} = 50$. We found from least-squares fits that $a=b=2.0$, $A=0.063$, and $B=11$. These coarse numerical approximations of $T'$ and $|\nabla T|_\mathrm{rms}$ as power-law functions of $Ma_\mathrm{t}$ only, along with the system of differential equations for the chemical kinetics mechanism, were then used inside a root-finding algorithm to determine the necessary values of $\varepsilon_\mathrm{inj}$, $L$, $T_\mathrm{init}$, and $P_\mathrm{init}$ for each simulation. 

Due to the high cost of simulating the chosen low-Mach conditions, only five simulations were performed, all targeted for $\xi_\mathrm{t}=4$, with one simulation at $\zeta_\mathrm{t}= 2$ and two simulations each at $\zeta_\mathrm{t} = 1$ and $0.5$, but with different $Ma_\mathrm{t}$ at the same $\zeta_\mathrm{t}$. Additionally, each simulation was initialized with a stoichiometric H$_2$-air mixture (i.e., H$_2$:O$_2$:N$_2=2$:$1$:$3.76$), and targeted a fixed value of $Re_\mathrm{t}=230$. Since we set four non-dimensional parameters for each simulation ($\xi_\mathrm{t}$, $\zeta_\mathrm{t}$, $Ma_\mathrm{t}$, and $Re_\mathrm{t}$), there is no remaining relational freedom to hold fixed any dimensional parameters between simulations, and, therefore, each simulation must have a different $T_\mathrm{init}$, $P_\mathrm{init}$, $L$, and $\varepsilon_\mathrm{inj}$. The input physical model parameters for each simulation and the resulting properties of the fully-developed turbulence are outlined in Table \ref{tab:setup}. It can be seen that the actual values of $\xi_\mathrm{t}$ and $\zeta_\mathrm{t}$, computed \emph{a posteriori} at $t=0$ in each case, maintain the trend of sweeping through unity $\zeta_\mathrm{t}$ at roughly constant $\xi_\mathrm{t}$, despite deviating from their precise target values. 

\begin{table*}[t!]
    \centering
    \begin{tabular}{r|ccccc}
    	\hline\hline
    	                                             Simulation &          Z1           &          Z2a          &          Z2b          &          Z3a          &          Z3b          \\
    	\hline
    	$\langle\rho\rangle\varepsilon_\mathrm{inj}$ (J/m$^3$s) & $5.556\times 10^{5}$  & $4.001\times 10^{4}$  & $1.524\times 10^{6}$  & $9.579\times 10^{4}$  & $2.066\times 10^{7}$  \\
    	                                                $L$ (m) & $3.278\times 10^{-2}$ & $1.116\times 10^{-1}$ & $3.087\times 10^{-2}$ & $1.113\times 10^{-1}$ & $1.729\times 10^{-1}$ \\
    	                                  $T_\mathrm{init}$ (K) &        $1118$         &        $998.1$        &        $1096$         &        $966.7$        &        $1119$         \\
    	                                $P_\mathrm{init}$ (atm) &        $6.987$        &        $1.796$        &        $4.585$        &        $1.096$        &        $4.160$        \\
    	\hline
    	                                    $Ma_{\mathrm{t}}$ &        $0.03$         &        $0.03$         &        $0.05$         &        $0.05$         &         $0.1$         \\
    	                                          $\{T\}_0$ (K) &        $1119$         &        $998.8$        &        $1098$         &        $968.5$        &        $1128$         \\
    	                            $\langle P \rangle_0$ (atm) &        $6.993$        &        $1.798$        &        $4.593$        &        $1.098$        &        $4.192$        \\
    	\hline
    	                             $T'_0$ (K) &       $0.1213$        &       $0.1068$        &       $0.2931$        &       $0.2648$        &        $1.183$        \\
    	                    $|\nabla T|_{\mathrm{rms},0}$ (K/m) &        $4.257$        &        $1.151$        &        $11.53$        &        $2.905$        &        $85.59$        \\
    	                                   $\lambda_{T,0}/\Delta x$ &         $8.9$         &         $8.5$         &         $8.5$         &         $8.4$         &         $8.2$         \\
    	\hline
    	                                     $\zeta_\mathrm{t}$ &        $2.87$         &        $1.47$         &        $1.45$         &        $0.69$         &        $0.67$         \\
    	                                       $\xi_\mathrm{t}$ &        $3.69$         &        $2.75$         &        $3.20$         &        $3.97$         &        $3.29$         \\
    	\hline\hline
    \end{tabular}
    \caption{Input physical model parameters of the direct numerical simulations and resulting actual turbulence properties, computed \emph{a posteriori}, at $t=0$~s.}
    \label{tab:setup}
\end{table*}

\begin{figure}[tb!]
    \centering
    \includegraphics[width=84mm]{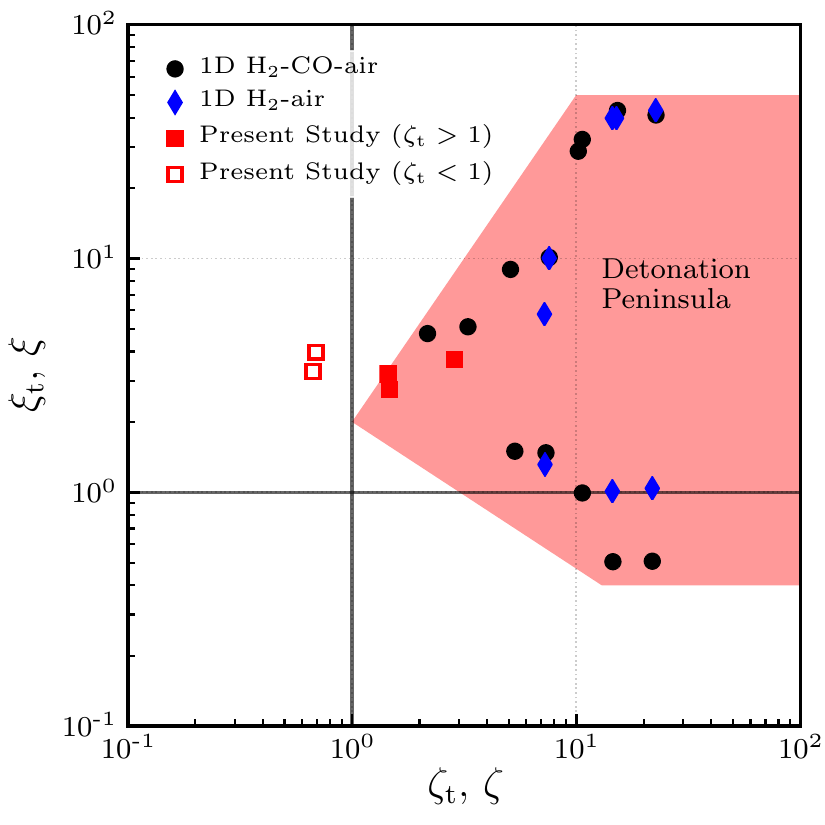}
    \caption{Hot-spot detonability regime diagram with present 3D turbulent simulation cases shown as red squares, with closed squares indicating $\zeta_\mathrm{t}>1$ and open squares indicating $\zeta_\mathrm{t}<1$. Also shown are select simulations of detonating 1D laminar hot spots of H$_2$-air (blue diamonds, \cite{Gu2003}) and H$_2$-CO-air (black circles, \cite{Bates2016}), which provide an empirical approximation of the upper and lower detonability limits. (color online)}
    \label{fig1}
\end{figure}

Figure \ref{fig1} shows a detonability regime diagram of the present simulations, along with a selection of detonating conditions from two previously published studies of 1D laminar hot spots containing H$_2$-air \cite{Gu2003} and H$_2$-CO-air \cite{Bates2016} fuel mixtures. Together, these prior studies sketch out the approximate upper and lower bounds of the hot-spot detonation peninsula. The 3D turbulent simulations presented here cut across the critical limit of $\zeta_\mathrm{t} = 1$, between the empirical upper and lower $\xi$ detonability limits. These simulation conditions therefore provide an adequate test of our hypothesis that the detonability of compressible turbulence should abruptly increase as $\zeta_\mathrm{t}$ increases from below to above unity. It is cautioned, however, that these simulations do not provide adequate information to draw firm conclusions about how the detonability may vary with either $\xi_\mathrm{t}$ or $Ma_\mathrm{t}$.

Each simulation was solved on a uniform computational mesh of size $1024^3$. The resulting Kolmogorov scale resolution at the time of reaction initiation was $\eta_{\textsc{k},0}/\Delta x = 0.5$, where $\Delta x$ is the grid cell size. This relatively coarse resolution was a necessary compromise between computational cost and physical fidelity, with the first four statistical moments of the temperature fluctuations being fully converged and the second moment of the temperature gradient field (i.e., $|\nabla T|_\mathrm{rms}$) being approximately converged, and with at least one grid cell per ZND detonation exothermic width for all simulations. Although the ZND detonation resolution is marginal, all characteristic gas-dynamic wave speeds, including reaction waves, are correctly captured by the Godunov finite-volume scheme, regardless of the grid size and wave orientation, and depend only on the Rankine-Hugoniot jump conditions across the wave.

We performed each simulation as follows. We randomly initialized each simulation with the chemical reactions turned off, uniform temperature and pressure, and a random velocity field with a prescribed isotropic energy spectrum and turbulence Mach number $Ma_\mathrm{t}$ significantly lower than the target turbulence Mach number. This initialization procedure was intended to ensure that any dynamical inconsistencies in the initial conditions did not impact the simulations at later times \cite{Ristorcelli1997, Samtaney2001}. The turbulent flow field was allowed to develop from these random initial conditions for $10.5\tau_\ell$, where $\tau_\ell$ is the predicted integral-scale eddy turnover time, until $Ma_\mathrm{t} = Ma_{\mathrm{t},0}$, and reactions were turned on. We define the time at which reactions are initialized as $t = 0$. We then ran the five simulations for an additional $1.0\tau_\ell$ each, at which point the entire domain of each simulation was essentially finished reacting. Full three-dimensional fields of the solution variables were output at a rate of $0.5\tau_\mathrm{exo}$. The turbulence forcing mechanism remains active throughout the entire duration of each simulation, and, due to the low values of $Ma_\mathrm{t}$ in all five cases, the thermodynamic turbulence statistics, $\{T\}$, $T'$, and $|\nabla T|_\mathrm{rms}$ would have been essentially stationary over a single $\tau_\ell$, were reactions not active. 

\begin{figure}[tb!]
    \centering
    \includegraphics[width=84mm]{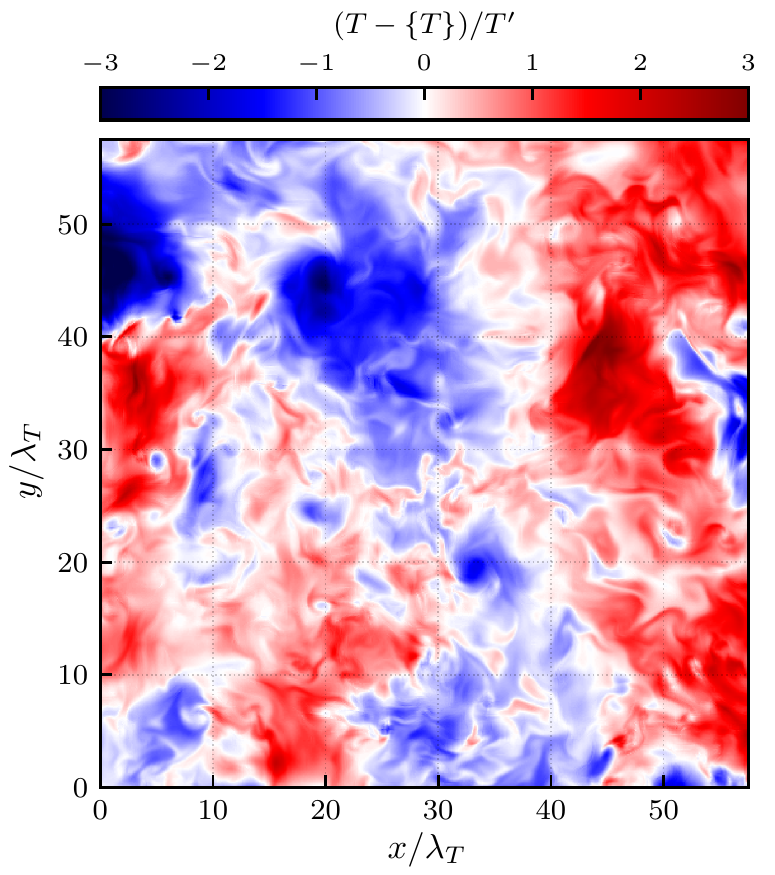}
    \caption{A pseudocolor image of a $512\times 512$ cell slice of the normalized temperature fluctuations, $(T - \{T\})/T'$, at $t=0$ from case Z1 (color online).}
    \label{fig2}
\end{figure}

A 2D section of the temperature field at $t=0$ is shown for case Z1 in Fig.~\ref{fig2}; this slice is qualitatively representative of the multi-scale structure of the temperature fluctuations for all five cases. Given the wide range of temperature values and associated temperature gradients found in Fig.~\ref{fig2}, the isotropic compressible turbulent flow fields generated in the present simulations are likely to form many isolated ignition kernels of various shapes and sizes. Therefore, the \emph{degree} to which a simulation forms detonation waves versus shockless auto-ignition waves can be quantified, rather than simply assessing a binary state of positive or negative detonability, as in the 1D laminar case. Even in cases Z3a and Z3b, where we do not predict that detonation waves will form since $\zeta_\mathrm{t}<1$ (see Fig.~\ref{fig1}), we still expect to find qualitative differences between combustion in an idealized homogeneous CV reactor and the spontaneous ignition waves that form along temperature gradients in these cases.

\subsection{Reaction wave profiling\label{ssec:profiles}}
\begin{figure}[tb!]
    \centering
    \includegraphics[width=78mm]{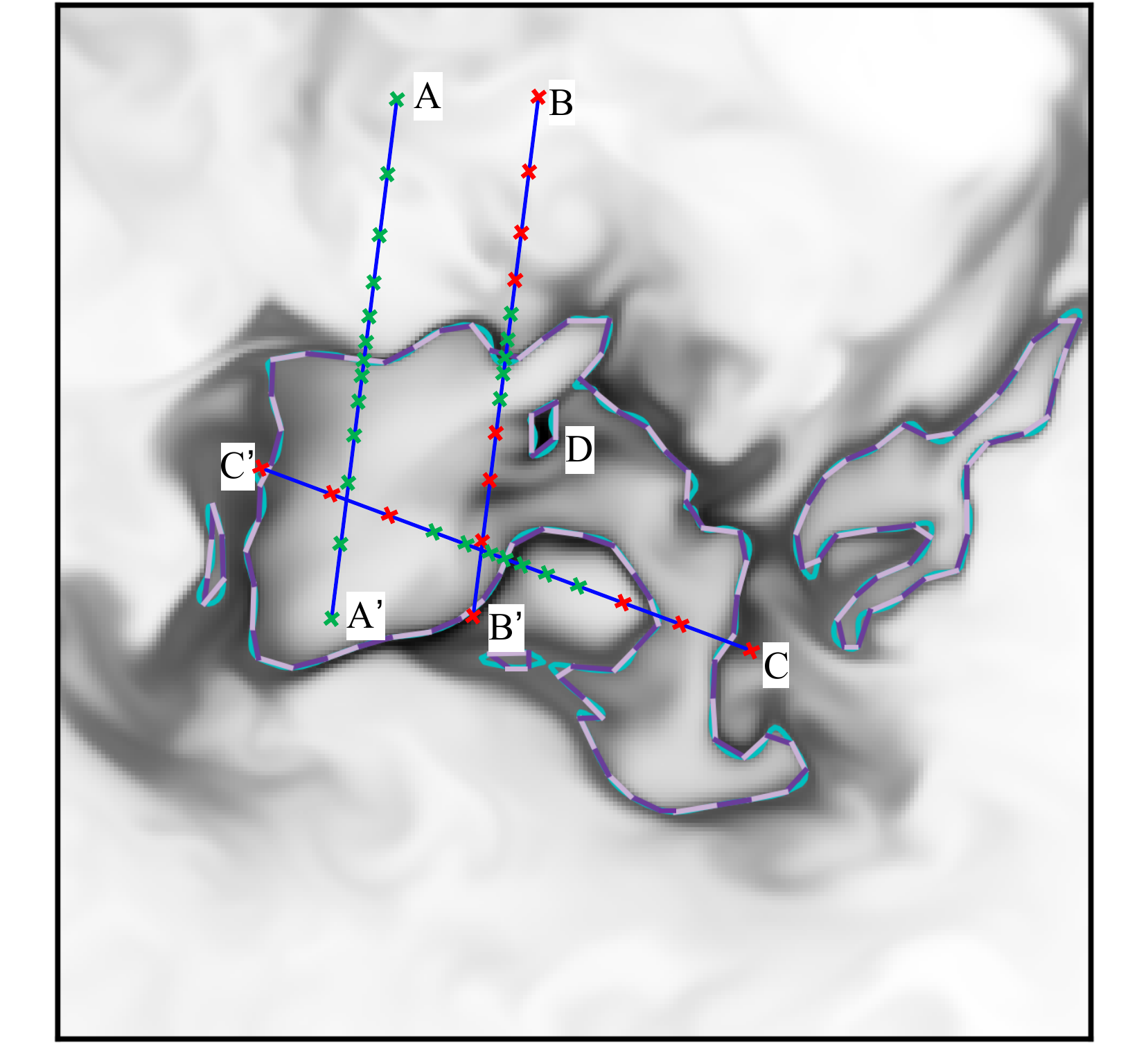}
    \caption{A two-dimensional representation of the reaction wave profiling procedure. See text for explanation. (color online)}
    \label{fig3}
\end{figure}

In Section \ref{sec:results}, we analyze the spontaneous ignition wave fronts within a single volumetric data output from each simulation. A 2D schematic representation of the reaction wave profiling is shown in Fig.~\ref{fig3} to aid explanation.

First, using Lewiner's marching cubes algorithm \cite{Lewiner2003}, as implemented in the python package scikit-image \cite{VanderWalt2014}, we find the isosurface of $Y_\mathrm{H_2}$ corresponding to the peak thermicity of a homogeneous CV reactor with the same initial conditions as the mean thermodynamic state of each simulation at $t=0$~s. The marching cubes algorithm builds a polygonal mesh made entirely of triangular elements whose vertices are the linearly-interpolated locations where the chosen value of a scalar field intersects each finite-volume cell edge. In Fig.~\ref{fig3}, the 3D piecewise-triangular isosurface is represented as a 2D piecewise-linear isocontour, shown as alternating light and dark purple lines, that approximates the ``true'' underlying ignition-delay isocontour, shown in cyan.

Second, for each triangular face of the isosurface, we compute the surface area, barycenter, and normal vector to the surface. From each face barycenter, we extract a reaction-front profile along the normal vector of the surface by interpolating all quantities of interest, computed at the finite-volume cell centers, to 29 geometrically-spaced points along each direction of the normal vector, such that each successive point is $2^{1/4}$ times farther from the face barycenter than the previous point. These points are located between $\pm 0.5\Delta x$ and $\pm 64\Delta x$, for a total of 59 interpolation points, including the face barycenter point, and quantities of interest are interpolated to these points using tri-cubic Akima spline interpolation \cite{Akima1970}. Three sample reaction-front profiles are represented in Fig.~\ref{fig3} as the blue lines marked $\overrightarrow{AA'}$, $\overrightarrow{BB'}$, and $\overrightarrow{CC'}$, with interpolation stencil points marked as green and red crosses. 

Third, we immediately truncate the interpolated profiles wherever they intersect another face of the isosurface, such as in the case of line $\overrightarrow{CC'}$ in Fig.~\ref{fig3}, which overlaps the 2D isocontour in three places other than the line segment from which it was interpolated. We also compute a smoothed thermicity profile using a five-point convolution filter and find the local maximum thermicity nearest the profile center-point. We then truncate the profiles to the nearest local minimum on either side of this center-most maximum. This step is demonstrated by displaying a smooth thermicity field as a gray-scale image underneath the 2D hydrogen mass-fraction isocontours in Fig.~\ref{fig3}. Any points which clearly cross over either another segment of the isocontour or a local minima in the thermicity field are colored red, indicating they have been discarded, and only the green points are retained. Any profile which has retained a complete reaction zone, defined as the region of exothermicity between the first and second half-max states, is considered to be a freely-propagating reaction wave profile, equivalent to a turbulent deflagration flamelet profile. 

It should be noted that interpolated profiles that lack a complete reaction zone may correspond to one of three scenarios: (i) \emph{Quasi-homogeneous autoignition}, wherein the underlying spontaneous autoignition wave would be longer than the periodic domain of the simulation (i.e., $\delta_\mathrm{rz} > L$); (ii) \emph{End-gas reactant pockets} bounded by colliding reaction waves, such that the thermicity has risen above 50\% of the profile peak everywhere along the profile within the pocket and therefore no first half-max point ($x_\mathrm{fhm}$) exists for the profile; or (iii) \emph{Emerging ignition kernels} bounded by cold reactants, such that the thermicity has not yet dropped below 50\% of the profile peak within the kernel and therefore no second half-max point ($x_\mathrm{shm}$) exists for the profile. An isolated end-gas reactant pocket marked as region $D$ in Fig.~\ref{fig3} is representative of sections of isosurface whose reaction-front profiles would not be retained due to the lack of a complete reaction zone structure.

\section{Results\label{sec:results}}

\subsection{Global system evolution}
\begin{figure*}[t!]
    \centering
    \includegraphics[width=164mm]{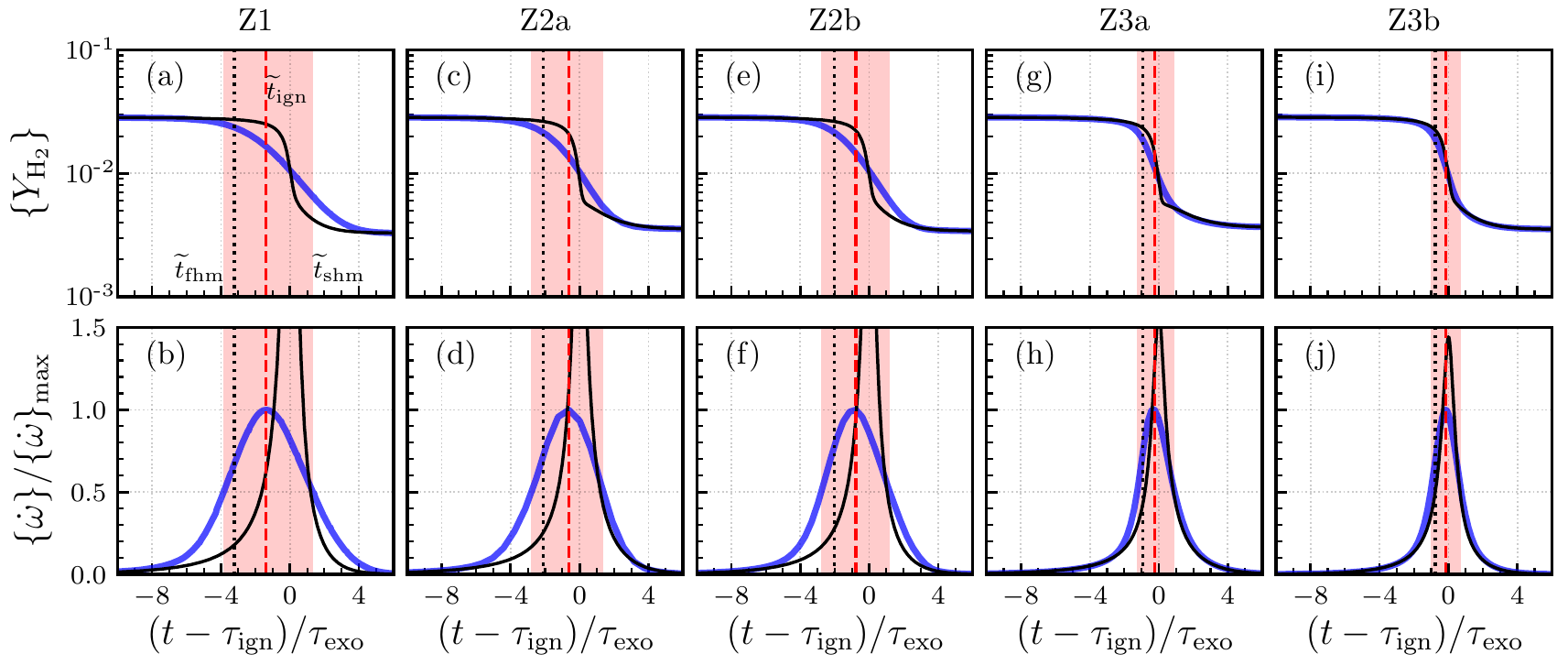}
    \caption{Time series of (a, c, e, g, i) mass-average H$_2$ mass-fraction, $\{Y_\mathrm{H_2}\}$, and (b, d, f, h, j) mass-average thermicity $\{\dot{\omega}\}$, shown as thick blue lines. Corresponding homogeneous autoignition profiles are shown as black lines, the domain-average ignition delay times, $\widetilde{t}_\mathrm{ign}$, are shown as vertical red lines, and the domain-average exothermicity times, $\widetilde{t}_\mathrm{exo}$, are shown as red shaded regions starting at $\widetilde{t}_\mathrm{fhm}$ and ending at $\widetilde{t}_\mathrm{shm}$. Vertical black dotted lines mark the time of the data output analyzed in each case. (color online)}
    \label{fig4}
\end{figure*}

Figure \ref{fig4} shows time series of the mass-average H$_2$ mass fraction, $\{Y_\mathrm{H_2}\}$, and thermicity, $\{\dot{\omega}\}$, for each turbulent simulation case. The corresponding time histories for a homogeneous CV reactor are also shown for the same spatially-averaged thermodynamic conditions found in each of the simulation cases at $t=0$~s (i.e., $\langle P\rangle_0$ and $\{T\}_0$). As $\zeta_\mathrm{t}$ decreases from case Z1 to Z3b, Fig.~\ref{fig4} shows that the rate at which $\{Y_\mathrm{H_2}\}$ changes in the simulations increasingly conforms to the rate of change in a homogeneous CV reactor. In particular, for case Z1, which has the largest value of $\zeta_\mathrm{t}$ of all simulations examined here, the rate of change of $\{Y_\mathrm{H_2}\}$ is substantially smaller than for the homogeneous CV reactor. Similarly, $\{\dot{\omega}\}$ is substantially smaller in the turbulent cases, as compared to the homogeneous CV reactor,  for large $\zeta_\mathrm{t}$ (i.e., case Z1). There is thus a close correspondence between the turbulent and homogeneous CV cases when $\zeta_\mathrm{t}$ is small and, as $\zeta_\mathrm{t}$ increases from case Z3b to Z1, the turbulent time histories increasingly diverge from those of the corresponding homogeneous CV reactors. 

Figure \ref{fig4} further shows that the global ignition delay time measured with respect to the $\{\dot{\omega}\}$ profiles, which we call the domain-average ignition delay time and denote as $\widetilde{t}_\mathrm{ign}$, decreases relative to the homogeneous CV ignition delay time, $\tau_\mathrm{ign}$, as $\zeta_\mathrm{t}$ increases. Figure \ref{fig4} also shows that there are substantial changes in the corresponding domain-averaged exothermicity time, $\widetilde{t}_\mathrm{exo} = \widetilde{t}_\mathrm{shm}-\widetilde{t}_\mathrm{fhm}$, relative to  $\tau_\mathrm{exo}$, as $\zeta_\mathrm{t}$ increases. In particular, $\widetilde{t}_\mathrm{exo}$ is more than five times the corresponding $\tau_\mathrm{exo}$ in case Z1, and still nearly double $\tau_\mathrm{exo}$ in case Z3b.

\begin{figure*}[t!]
    \centering
    \includegraphics[width=164mm]{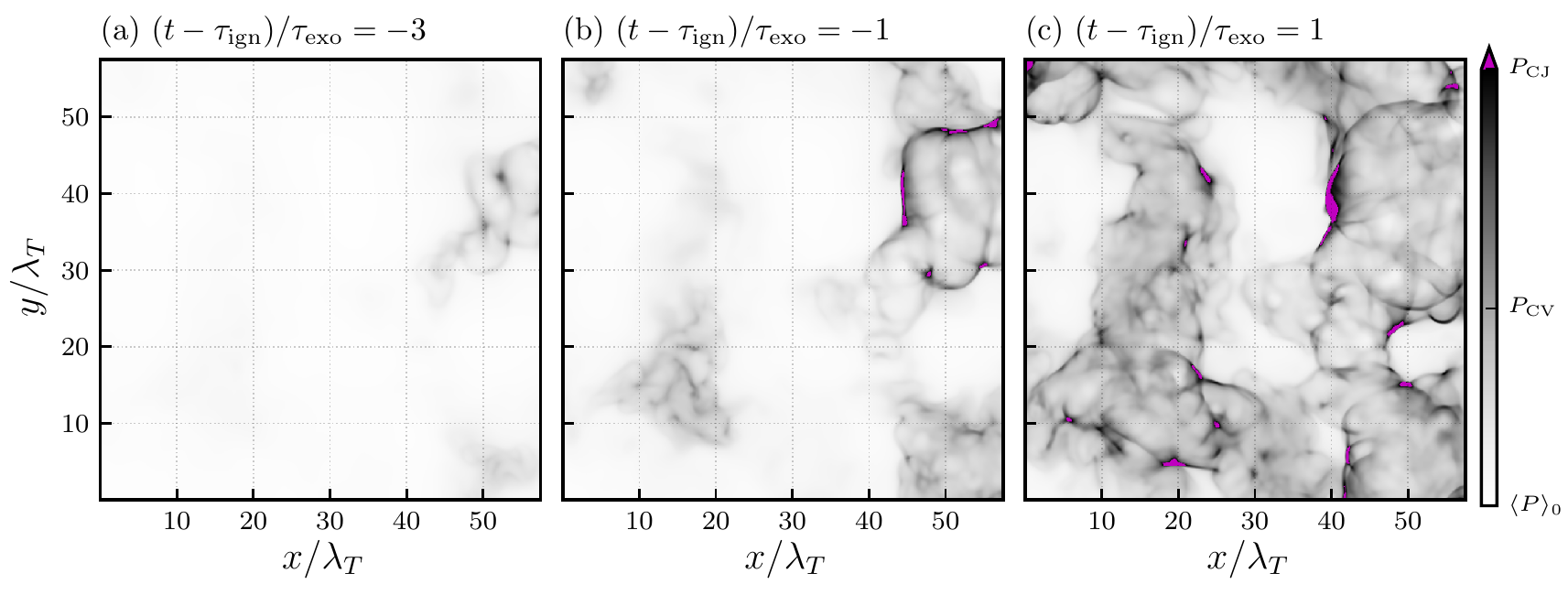}
    \caption{Pseudocolor images of $512\times 512$ cell slices of the pressure field, $P$, taken from case Z1 at (a) $(t-\tau_\mathrm{ign})/\tau_\mathrm{exo} = -3$, (b) $(t-\tau_\mathrm{ign})/\tau_\mathrm{exo} = -1$, and (c) $(t-\tau_\mathrm{ign})/\tau_\mathrm{exo} = 1$, approximately corresponding to $\widetilde{t}_\mathrm{fhm}$, $\widetilde{t}_\mathrm{ign}$, and $\widetilde{t}_\mathrm{shm}$. Pressures greater than $P_\mathrm{CJ}$ are highlighted as magenta. (color online)}
    \label{fig5}
\end{figure*}

The spatial structure of the simulations during the domain-averaged exothermic pulse is indicated by the partial 2D slices of pressure at three times shown in Fig.~\ref{fig5} for case Z1. The initial emergence of isolated ignition kernels just before the first domain-average half-max point, $\widetilde{t}_\mathrm{fhm}$, is shown in Fig.~\ref{fig5}(a), while Fig.~\ref{fig5}(b) shows that reaction waves from the ignition kernels have rapidly strengthened and are interacting with secondary ignition kernels just after the domain-average ignition-delay time, $\widetilde{t}_\mathrm{ign}$. By the second domain-average half-max point, $\widetilde{t}_\mathrm{shm}$, the flow field is dominated by colliding reaction waves and merging ignition kernels, as shown in Fig.~\ref{fig5}(c). Therefore, to specifically examine hot-spot autoignition and detonation formation, in the following we analyze the single volumetric data output from each simulation that most closely corresponds to $\widetilde{t}_\mathrm{fhm}$, marked by vertical black dotted lines in Fig.~\ref{fig4} and shown for case Z1 in Fig.~\ref{fig5}(a). It should be noted that while the analysis of the ignition front isosurfaces and their associated reaction-front profiles is performed at $t=\widetilde{t}_\mathrm{fhm}$, each simulation continues to be characterized by the values of $\xi_\mathrm{t}$ and $\zeta_\mathrm{t}$ computed at $t=0$.

\subsection{Reaction wave classification}
The simple configuration of an isolated laminar hot-spot ensures that there can only be self-reinforcing acoustic-exothermicity coupling, as both the exothermic ignition wave and acoustic wave formed from initial conditions travel in the same direction along a negative gradient in the speed of sound. Similarly, once the reaction wave has exited the hot spot, it is assumed to travel through a quiescent cold gas at a fixed wave speed. Accordingly, the expected reaction wave speed inside the hot spot (i.e., $\xi$) is a critical indicator of detonability, and the final steady-state wave speed alone could be used as an indicator of successful detonation. 

However, in the case of igniting compressible turbulence, it is possible that, locally and instantaneously, an exothermic reaction wave may either support or oppose an acoustic or shock wave, depending on the degree of misalignment between the waves, as compressible turbulence hot spots are not isolated from one another, do not form in a quiescent acoustic field, and the temperature and speed of sound do not monotonically decrease away from the hot spot. Instead, the local and instantaneous wave speed along a reaction wave profile, which could be computed as the ignition isosurface displacement speed as in \cite{Chen2006}, can be expected to vary rapidly and significantly in time, because the ignition-delay-time gradient along the direction of the reaction wave profile will tend to oscillate above and below zero, and the spontaneous wave speed will therefore tend to intermittently shoot to infinity along the profile.

For this reason, we do not classify reaction wave profiles based upon their instantaneous isosurface displacement speed. Instead, we classify each reaction wave identified in the five simulations based on both the magnitude of the profile-maximum pressure, $P_\mathrm{max}$, which we take as a measure of the magnitude of the acoustic-exothermicity coupling, and the relative position of the pressure peak with respect to the profile reaction zone, which we take as a measure of the direction of the coupling between the reaction and pressure wave. We delineate detonation waves of various strengths from ignition waves with marginal or weak acoustic coupling by comparing $P_\mathrm{max}$ to the post-shock pressure for a planar ZND detonation wave propagating through a gas at the mean thermodynamic state of each simulation at $t=0$~s, denoted $P_\mathrm{ZND}$, the corresponding CJ post-detonation sonic-point pressure of that ZND detonation, $P_\mathrm{CJ}$, and the final pressure of a homogeneous CV reactor with the same initial conditions, $P_\mathrm{CV}$. These criteria are:
\begin{itemize}[itemsep=0pt,leftmargin=*]
    \item If $P_\mathrm{max} \ge P_\mathrm{ZND}$, then the reaction wave is a \emph{high-pressure detonation};
    \item If $P_\mathrm{ZND} > P_\mathrm{max} \ge P_\mathrm{CJ}$, then the reaction wave is a \emph{low-pressure detonation};
    \item If $P_\mathrm{CJ} > P_\mathrm{max} \ge 1.1P_\mathrm{CV}$, then the reaction wave is a \emph{high-pressure ignition}; finally,
    \item If $P_\mathrm{max} < 1.1P_\mathrm{CV}$, then the reaction wave is a \emph{low-pressure ignition}.
\end{itemize}

\begin{figure*}[b!]
    \centering
    \includegraphics[width=164mm]{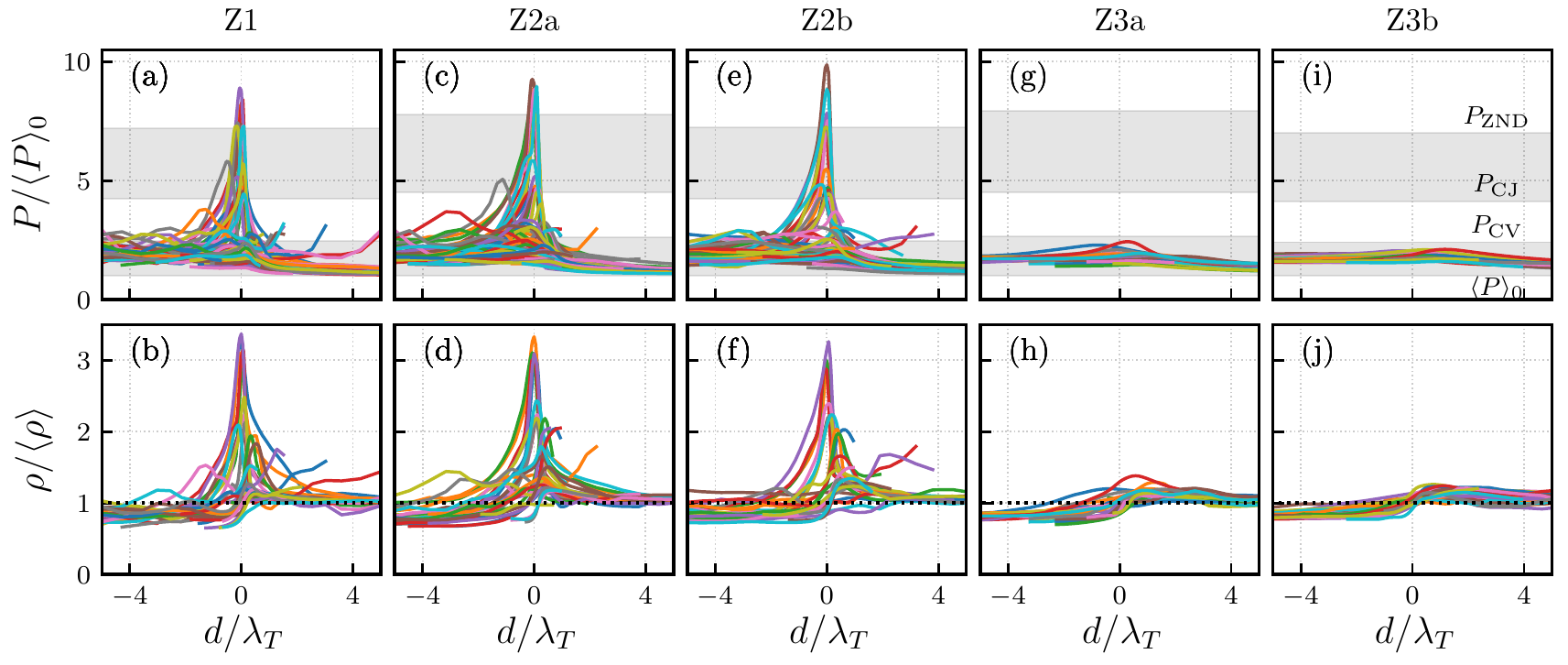}
    \caption{Forty selected reaction wave profiles of (a, c, e, g, i) pressure, $P$, and (b, d, f, h, j) density, $\rho$, shown as solid colored lines. Distance along the profile interpolation stencils is denoted as $d$, with $d=0$ corresponding to the isosurface face barycenter, positive $d$ on the upstream (reactant) side of the reaction wave, and negative $d$ on the downstream (product) side of the reaction wave. The various pressure regimes used for wave classification are denoted by alternating bands of gray shading in panels (a)-(i). (color online).}
    \label{fig6}
\end{figure*}

\begin{figure}[tb!]
    \centering
    \includegraphics[width=84mm]{{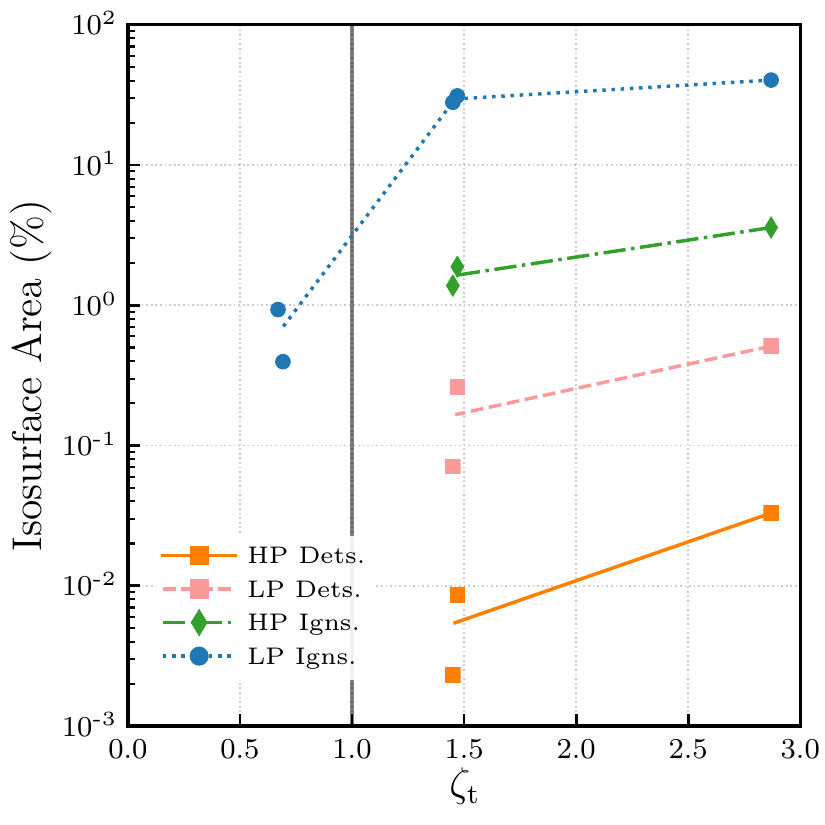}}
    \caption{Percentage of total ignition-isosurface area corresponding to each reaction wave classification as a function of $\zeta_\mathrm{t}$, directly computed at $t=0$ for each of the five simulation cases. Colors and symbols correspond to HP detonation waves (orange squares), LP detonation waves (pink squares), HP ignition waves (green diamonds), and LP ignition waves (blue circles). Piecewise-linear trend lines are shown by averaging the area fractions of cases Z2a with Z2b, and Z3a with Z3b. (color online)}
    \label{fig7}
\end{figure}

Reaction front profiling resulted in several million individual profiles in each simulation, and so we show a selected small number of the resulting reaction wave profiles for each case in Fig.~\ref{fig6}. Based on the above criteria, we find detonation waves in all three cases where $\zeta_\mathrm{t} > 1$ (namely, Z1, Z2a, and Z2b) and no detonations in cases Z3a and Z3b. Figure \ref{fig6} shows a subset of the reaction wave profiles identified in each of the simulations. Both high-pressure (HP) and low-pressure (LP) detonations are observed in the detonating cases, although only a very small fraction of the total ignition isosurface area is classified as either a detonation wave or HP ignition wave, as shown in Fig.~\ref{fig7}. In cases Z3a and Z3b, only a very small fraction of the total isosurface area is categorized as a complete reaction wave at all, with all identified waves being LP ignition waves. It should be noted that the area fraction of the ignition isosurfaces associated with reaction profiles that lack a complete reaction zone are not plotted in Fig.~\ref{fig7} and the sum of the reaction-wave area fractions is less than 100\% in all five simulation cases.

The position of maximum pressure in each profile relative to its reaction zone allows for a more refined assessment of the acoustic-exothermicity coupling of each detonation wave, and therefore for the system as a whole. In particular, we can sub-classify the stability of HP and LP detonations according to the position of the maximum pressure, $x_\mathrm{mp} = x|_{P = P_\mathrm{max}}$. Specifically, if $x_\mathrm{mp} > x_\mathrm{fhm}$, then the reaction zone is behind a leading shock and we consider this to be a \emph{positively-coupled} detonation. If $x_\mathrm{fhm} \ge x_\mathrm{mp} \ge x_\mathrm{shm}$, then the reaction zone overlaps the shock and we consider this to be a \emph{neutrally-coupled} detonation. Finally, if $x_\mathrm{shm} > x_\mathrm{mp}$, then the reaction zone is in front of a trailing shock and we consider this to be a \emph{negatively-coupled} detonation.

We found no positively-coupled HP detonations in the three $\zeta_\mathrm{t} > 1$ cases and classify less than $0.001\%$ of the isosurface area as a positively-coupled LP detonation in each of those cases. Moreover, we classify just $0.0330\%$, $0.0086\%$, and $0.0023\%$ of the total ignition surface as neutrally-coupled HP detonations in cases Z1, Z2a, and Z2b, respectively.

We note that the marginally-interacting HP ignition waves (i.e., HPI waves), could also be sub-classified in a similar manner, and in fact it is possible that a wave classified as an HPI wave is in fact an emerging or very LP detonation wave. However, due to the nature of the single-time Eulerian analysis we use here, it is impossible to properly assess whether or not a given HPI wave will eventually develop into a detonation wave.

\subsection{Evolution in thermodynamic state space}

\begin{figure}[tbph!]
    \centering
    \includegraphics[width=84mm]{{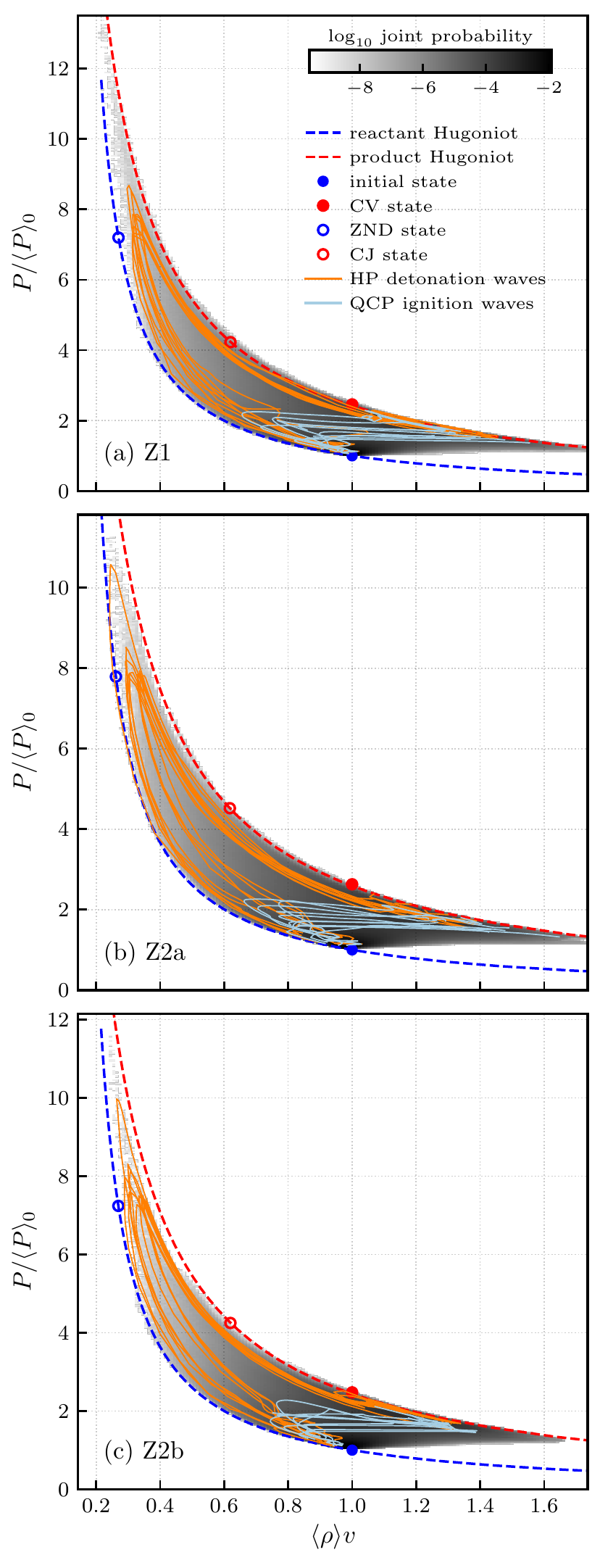}}
    \caption{Joint probability distributions of pressure, $P$, and specific volume, $v$, over the full domains of cases (a) Z1, (b) Z2a, and (c) Z2b, along with 10 randomly-chosen examples each of developing HP detonations (thin orange lines) and quasi-constant-pressure ignition waves (thin light blue lines). Lines, symbols, and colors are defined for all cases in panel (a) (color online).}
    \label{fig8}
\end{figure}

\begin{figure}[t!]
    \centering
    \includegraphics[width=84mm]{{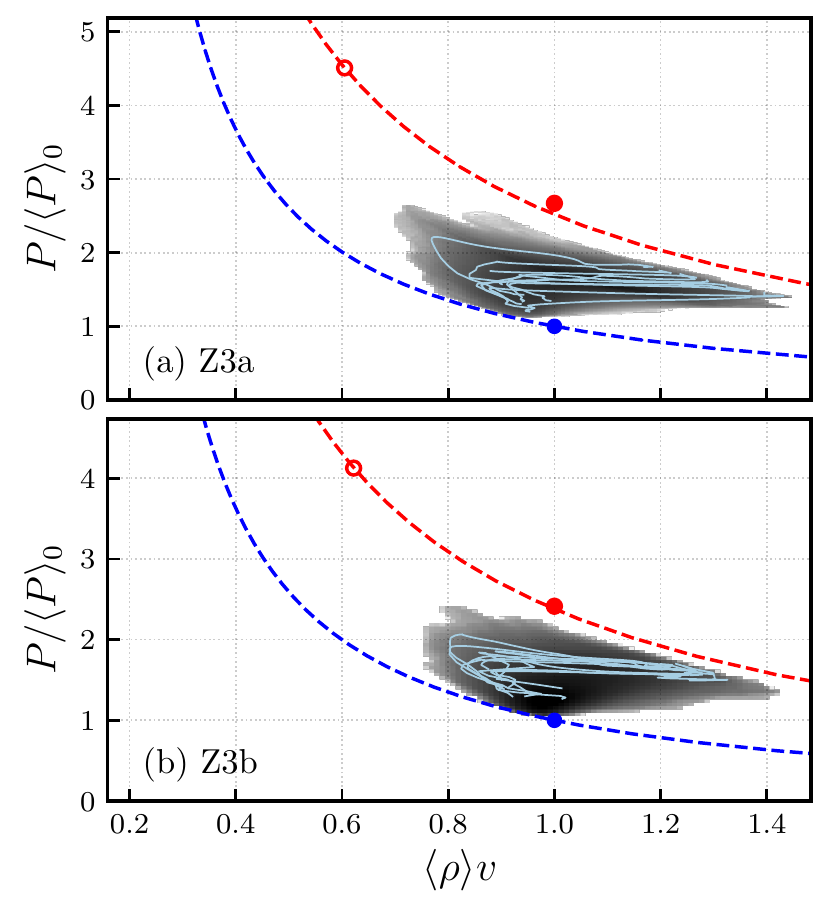}}
    \caption{Joint probability distributions of pressure, $P$, and specific volume, $v$, over the full domains of cases (a) Z3a, and (b) Z3b, along with 10 randomly-chosen examples of quasi-constant-pressure ignition waves. Lines and symbols as in Fig.~\ref{fig8}(a) (color online).}
    \label{fig9}
\end{figure}

In order to directly illustrate the variety of thermodynamic cycles that each reacting fluid parcel may undergo locally and instantaneously throughout the domain, including well-defined, if less than ideal, detonation wave cycles, joint probability distributions (pdfs) of pressure $P$ versus specific volume $v = 1/\rho$ are constructed from the full domain of volumetric data that was analyzed in each case. Figure \ref{fig8} shows the variation in the joint pdf of the $P$--$v$ state space for the three cases (Z1, Z2a, and Z2b) that exhibit strong acoustic-exothermic coupling, including detonations, while Fig.~\ref{fig9} shows the two cases (Z3a and Z3b) with very weak acoustic-exothermicity coupling throughout their domains.

In each panel of Figs.~\ref{fig8} and \ref{fig9}, the joint pdf has an envelope corresponding to the chemically-frozen Hugoniot curve for pure reactants (dashed blue lines) and the chemical-equilibrium Hugoniot curve of the reaction products (dashed red lines). Each panel also highlights the average initial thermodynamic state at $t = 0$ s, ($\langle P\rangle_0$, $1/\langle\rho\rangle$), the von Neumann post-shock state, ($P_\mathrm{ZND}$, $v_\mathrm{ZND}$), Chapman-Jouguet post-detonation sonic state, ($P_\mathrm{CJ}$, $v_\mathrm{CJ}$), and the equilibrium state for a homogeneous constant-volume reactor, ($P_\mathrm{CV}$, $v_\mathrm{CV}$). Of particular note, Fig.~\ref{fig9} reveals that in neither case Z3a nor Z3b has any point within the domain fully reacted, since, in both cases, the joint pdf falls noticeably short of the equilibrium product Hugoniot curve.

For the reaction waves which show only very weak acoustic-exothermicity coupling, namely the LPI waves, there is generally no sharp pressure peak [see, for instance, panels (g) and (i) of Fig.~\ref{fig6}], and therefore it is easier to sub-classify these LPI waves by assessing whether or not they follow an approximately constant pressure or constant volume trajectory through state space. Therefore, we compute the normalized ratio of the change in pressure over the change in specific volume, $v$, across the reaction zone of a wave as 
\begin{equation}
m = \frac{v_\mathrm{fhm}|P_\mathrm{shm} - P_\mathrm{fhm}|}{P_\mathrm{fhm}|v_\mathrm{shm} - v_\mathrm{fhm}|}\,.
\end{equation}
If $m > 4$, then we consider the LPI to be a quasi-constant-volume ignition wave (QCV), and if $m < 1/4$, then the LPI wave is a quasi-constant-pressure ignition wave (QCP). We chose the threshold slopes of $4$ and $1/4$ empirically, based upon trial and error, until we found reciprocal slopes that adequately isolated waves with relatively straight (whether horizontal or vertical) state-space trajectories within the reaction zone of each profile. We found a significant fraction of the LPI waves in all five cases undergo QCP heat release and both Figs.~\ref{fig8} and \ref{fig9} show ten examples of QCP waves, randomly-chosen from among all identified QCP profiles in each case. For comparison, Fig.~\ref{fig8} also shows ten randomly-chosen strong developing detonation wave profiles for cases Z1, Z2a, and Z2b. 

\section{Conclusions}
In autoignitive gaseous flows, detonations can be directly initiated by the thermomechanical feedback between spontaneous ignition waves and acoustic waves which emanate from isolated and laminar hot spots with suitable temperature gradients. Two non-dimensional parameters have been found to predict the detonability of such hot spots, namely the acoustic-ignition coupling parameter, $\xi$, which is an inverse Mach number for the ignition wave, and the acoustic-exothermicity coupling parameter, $\zeta$, which is a ratio of the sound-crossing time to the exothermic time of the hot spot. In this work, we have adapted the definitions of $\xi$ and $\zeta$ into a statistical model for spontaneous detonation initiation by thermodynamic gradients formed from compressible homogeneous isotropic turbulence (HIT) fluctuations. We then examined the utility of the adapted statistical model for predicting turbulence detonability in five new direct numerical simulations (DNS) of compressible HIT in premixed hydrogen-air that sweep a small, but critical, range of the turbulence detonability parameters, namely at values of $\zeta_\mathrm{t}$ both above and below unity. 

We found strong evidence that $\zeta_\mathrm{t}$ is a sensitive measure of the degree of acoustic-exothermicity coupling between the turbulence thermodynamics and the combustion chemistry, as all three simulation cases where $\zeta_\mathrm{t} > 1$ contained developing detonation waves, and the two simulation cases where $\zeta_\mathrm{t} < 1$ showed no evidence of strong coupling between spontaneous ignition waves and the very weak leading pressure waves formed as isolated ignition kernels emerged. Additionally, while detonation waves accounted for only a very small fraction of the total ignition-front isosurface area in each case where $\zeta_\mathrm{t} > 1$, the area fraction of detonation, detonation strength, and detonation stability all increased monotonically with $\zeta_\mathrm{t}$. We conclude, therefore, that our formulation of $\zeta_\mathrm{t}$ is a predictive parameter of compressible turbulence detonability.

There are, however, several limitations of the present study, which limit its immediate applicability to the engineering design process. Namely, the choice of HIT as the unit flow problem simplified the required dimensional analysis but greatly limits the ability to extrapolate these findings to the prediction of detonability of geometrically-complex turbulent flows with large-scale variations in autoignitive conditions, as would be found in any real-world application. Similarly, the limited range of turbulence Mach and Reynolds numbers, $Ma_\mathrm{t}$ and $Re_\mathrm{t}$, that were computationally accessible for this study precluded testing whether these parameters also have a direct impact on detonability, separate from their influence on $\zeta_\mathrm{t}$. Additionally, due to the approximate nature of the derived scaling of $|\nabla T|_\mathrm{rms}$ with $Ma_\mathrm{t}$, $Re_\mathrm{t}$, $Pr$, and $\gamma$, we are unable to draw any definitive conclusions about how detonability varies with $\xi_\mathrm{t}$ from the paired simulations Z2a/Z2b and Z3a/Z3b. However, the high level of consistency of the results observed between the two pairs of cases provides confidence in our ability to predict detonability based on $\xi_\mathrm{t}$ and $\zeta_\mathrm{t}$ alone, independent of $Ma_\mathrm{t}$ and the thermodynamic properties of a system. High computational cost also forced us to choose between simulating an interesting set of points in the $\xi_\mathrm{t}$--$\zeta_\mathrm{t}$ parameter space and performing simulations on highly-refined meshes. As a result, given the diffusive, monotone, and shock-capturing nature of the numerics, it is likely that each case would exhibit a higher degree of detonability were the simulations run with a much higher mesh resolution capable of resolving the internal thermochemical structure of the detonation waves.

For these reasons, future work will focus on empirically determining the non-dimensional scaling of $T'$ and $|\nabla T|_\mathrm{rms}$ with $Ma_\mathrm{t}$, $Re_\mathrm{t}$, $Pr$, and $\gamma$, performing detonability simulations at both higher $Ma_\mathrm{t}$ and $Re_\mathrm{t}$, and determining how superimposed compressible turbulence fluctuations alter the detonability of large-scale and coherent temperature gradients. Future work could also test alternative statistical formulations of the turbulence detonability model that use a model for the isotropic temperature autocorrelation function, or other forms of higher-order statistical information, to better capture the contribution of the highest temperature regions of the domain and the largest temperature gradients.

\section*{Acknowledgments}
The authors would like to thank D.~R.~Kassoy for helpful discussions. CAZT and PEH were supported in part by AFOSR awards FA9550-14-1-0273 and FA9550-17-1-0144, and by the Department of Defense (DoD) High Performance Computing Modernization Program (HPCMP) under a Frontier Project award. Computing resources were provided by the DoD HPCMP under the Frontier project award.

\bibliographystyle{elsarticle-num.bst}

\begin{thebibliography}{10}
    \expandafter\ifx\csname url\endcsname\relax
    \def\url#1{\texttt{#1}}\fi
    \expandafter\ifx\csname urlprefix\endcsname\relax\def\urlprefix{URL }\fi
    \expandafter\ifx\csname href\endcsname\relax
    \def\href#1#2{#2} \def\path#1{#1}\fi
    
    \bibitem{Urzay2018}
    J.~Urzay, Supersonic combustion in air-breathing propulsion systems for
    hypersonic flight, Annual Review of Fluid Mechanics 50 (2018) 593--627.
    
    \bibitem{Oran2007}
    E.~S. Oran, V.~N. Gamezo, Origins of the deflagration-to-detonation transition
    in gas-phase combustion, Combustion and Flame 148 (2007) 4--47.
    
    \bibitem{Zeldovich1970}
    Y.~B. Zel'dovich, V.~B. Librovich, G.~M. Makhviladze, G.~I. Sivashinsky, On the
    development of detonation in a non-uniformly heated gas, Astronautica Acta 15
    (1970) 313--321.
    
    \bibitem{Lee1978}
    J.~H. Lee, R.~Knystautas, N.~Yoshikawa, Photochemical initiation of gaseous
    detonations, Acta Astronautica 5~(11-12) (1978) 971--982.
    
    \bibitem{Zeldovich1980}
    Y.~B. Zel'dovich, Regime classification of an exothermic reaction with
    nonuniform initial conditions, Combustion and Flame 39~(2) (1980) 211--214.
    
    \bibitem{Bradley1996}
    D.~Bradley, ‘hot spots’ and gasoline engine knock, Journal of the Chemical
    Society, Faraday Transactions 92~(16) (1996) 2959--2964.
    
    \bibitem{Khokhlov1991}
    A.~Khokhlov, Mechanisms for the initiation of detonations in the degenerate
    matter of supernovae, Astronomy and Astrophysics 246 (1991) 383--396.
    
    \bibitem{Khokhlov1997}
    A.~M. Khokhlov, E.~S. Oran, J.~C. Wheeler, A theory of
    deflagration-to-detonation transition in unconfined flames, Combustion and
    Flame 108~(4) (1997) 503--517.
    
    \bibitem{Poludnenko2011a}
    A.~Y. Poludnenko, T.~A. Gardiner, E.~S. Oran, {Spontaneous transition of
        turbulent flames to detonations in unconfined media}, Physical Review Letters
    107~(5) (2011) 1--4.
    
    \bibitem{Poludnenko2015}
    A.~Y. Poludnenko, {Pulsating instability and self-acceleration of fast
        turbulent flames}, Physics of Fluids 27~(1) (2015) 014106.
    
    \bibitem{Kassoy2010}
    D.~R. Kassoy, The response of a compressible gas to extremely rapid transient,
    spatially resolved energy addition: an asymptotic formulation, Journal of
    Engineering Mathematics 68~(3-4) (2010) 249--262.
    
    \bibitem{Bartenev2000}
    A.~M. Bartenev, B.~E. Gelfand, Spontaneous initiation of detonations, Progress
    in Energy and Combustion Science 26~(1) (2000) 29--55.
    
    \bibitem{Kapila2002}
    A.~K. Kapila, D.~W. Schwendeman, J.~J. Quirk, T.~Hawa, Mechanisms of detonation
    formation due to a temperature gradient, Combustion Theory and Modelling
    6~(4) (2002) 553--594.
    
    \bibitem{Gu2003}
    X.~J. Gu, D.~R. Emerson, D.~Bradley, Modes of reaction front propagation from
    hot spots, Combustion and Flame 133~(1) (2003) 63--74.
    
    \bibitem{Radulescu2013}
    M.~I. Radulescu, G.~J. Sharpe, D.~Bradley, A universal parameter quantifying
    explosion hazards, detonability and hot spot formation: $\chi$ number, in:
    Fire and Explosion Hazards: Proceedings of the Seventh International Seminar,
    2013.
    
    \bibitem{Regele2013}
    J.~D. Regele, D.~R. Kassoy, A.~Vezolainen, O.~V. Vasilyev, Indirect detonation
    initiation using acoustic timescale thermal power deposition, Physics of
    Fluids 25~(9) (2013) 091113.
    
    \bibitem{Kurtz2014}
    M.~D. Kurtz, J.~D. Regele, Acoustic timescale characterisation of a
    one-dimensional model hot spot, Combustion Theory and Modelling 18~(4-5)
    (2014) 532--551.
    
    \bibitem{Bates2016}
    L.~Bates, D.~Bradley, G.~Paczko, N.~Peters, Engine hot spots: Modes of
    auto-ignition and reaction propagation, Combustion and Flame 166 (2016)
    80--85.
    
    \bibitem{Frisch1995}
    U.~Frisch, Turbulence: the Legacy of AN Kolmogorov, Cambridge University Press,
    1995.
    
    \bibitem{Sagaut2008}
    P.~Sagaut, C.~Cambon, Homogeneous Turbulence Dynamics, Cambridge University
    Press, 2008.
    
    \bibitem{Schumacher2014}
    J.~Schumacher, J.~D. Scheel, D.~Krasnov, D.~A. Donzis, V.~Yakhot, K.~R.
    Sreenivasan, Small-scale universality in fluid turbulence, Proceedings of the
    National Academy of Sciences 111~(30) (2014) 10961--10965.
    
    \bibitem{Poludnenko2010}
    A.~Y. Poludnenko, E.~S. Oran, {The interaction of high-speed turbulence with
        flames: Global properties and internal flame structure}, Combustion and Flame
    157~(5) (2010) 995--1011.
    \newblock \href {http://arxiv.org/abs/1106.3699} {\path{arXiv:1106.3699}}.
    
    \bibitem{Hamlington2011a}
    P.~E. Hamlington, A.~Y. Poludnenko, E.~S. Oran, {Interactions between
        turbulence and flames in premixed reacting flows}, Physics of Fluids 23~(12).
    
    \bibitem{Aspden2011}
    A.~J. Aspden, M.~S. Day, J.~B. Bell, Turbulence--flame interactions in lean
    premixed hydrogen: transition to the distributed burning regime, Journal of
    Fluid Mechanics 680 (2011) 287--320.
    
    \bibitem{Hawkes2012}
    E.~R. Hawkes, O.~Chatakonda, H.~Kolla, A.~R. Kerstein, J.~H. Chen, A petascale
    direct numerical simulation study of the modelling of flame wrinkling for
    large-eddy simulations in intense turbulence, Combustion and Flame 159~(8)
    (2012) 2690--2703.
    
    \bibitem{Kolla2014}
    H.~Kolla, E.~R. Hawkes, A.~R. Kerstein, N.~Swaminathan, J.~H. Chen, On velocity
    and reactive scalar spectra in turbulent premixed flames, Journal of Fluid
    Mechanics 754 (2014) 456--487.
    
    \bibitem{Towery2016}
    C.~A.~Z. Towery, A.~Y. Poludnenko, J.~Urzay, J.~O'Brien, M.~Ihme, P.~E.
    Hamlington, Spectral kinetic energy transfer in turbulent premixed reacting
    flows, Physical Review E 93~(5) (2016) 053115.
    
    \bibitem{Obrien2017}
    J.~O'Brien, C.~A.~Z. Towery, P.~E. Hamlington, M.~Ihme, A.~Y. Poludnenko,
    J.~Urzay, {The cross-scale physical-space transfer of kinetic energy in
        turbulent premixed flames}, Proceedings of the Combustion Institute 36~(2)
    (2017) 1967---1975.
    
    \bibitem{Kim2018}
    J.~Kim, M.~Bassenne, C.~A.~Z. Towery, P.~E. Hamlington, A.~Y. Poludnenko,
    J.~Urzay, Spatially-localized multi-scale energy transfer in turbulent
    premixed combustion, Journal of Fluid Mechanics 848 (2018) 78--116.
    
    \bibitem{Whitman2019}
    S.~H.~R. Whitman, P.~E. Hamlington, C.~A.~Z. Towery, A.~Y. Poludnenko, Scaling
    and collapse of conditional velocity structure functions in turbulent
    premixed flames, Proceedings of the Combustion Institute 37~(2) (2019)
    2527--2535.
    
    \bibitem{Sankaran2005}
    R.~Sankaran, H.~G. Im, E.~R. Hawkes, J.~H. Chen, The effects of non-uniform
    temperature distribution on the ignition of a lean homogeneous hydrogen--air
    mixture, Proceedings of the Combustion Institute 30~(1) (2005) 875--882.
    
    \bibitem{Chen2006}
    J.~H. Chen, E.~R. Hawkes, R.~Sankaran, S.~D. Mason, H.~G. Im, Direct numerical
    simulation of ignition front propagation in a constant volume with
    temperature inhomogeneities: I. fundamental analysis and diagnostics,
    Combustion and Flame 145~(1) (2006) 128--144.
    
    \bibitem{Hawkes2006}
    E.~R. Hawkes, R.~Sankaran, P.~P. P{\'e}bay, J.~H. Chen, Direct numerical
    simulation of ignition front propagation in a constant volume with
    temperature inhomogeneities: {II.} parametric study, Combustion and Flame
    145~(1) (2006) 145--159.
    
    \bibitem{Yoo2011}
    C.~S. Yoo, T.~Lu, J.~H. Chen, C.~K. Law, Direct numerical simulations of
    ignition of a lean n-heptane/air mixture with temperature inhomogeneities at
    constant volume: Parametric study, Combustion and Flame 158~(9) (2011)
    1727--1741.
    
    \bibitem{Im2015}
    H.~G. Im, P.~Pal, M.~S. Wooldridge, A.~B. Mansfield, A regime diagram for
    autoignition of homogeneous reactant mixtures with turbulent velocity and
    temperature fluctuations, Combustion Science and Technology 187~(8) (2015)
    1263--1275.
    
    \bibitem{Liberman2012}
    M.~A. Liberman, A.~D. Kiverin, M.~F. Ivanov, Regimes of chemical reaction waves
    initiated by nonuniform initial conditions for detailed chemical reaction
    models, Physical Review E 85~(5) (2012) 056312.
    
    \bibitem{Dai2015}
    P.~Dai, Z.~Chen, S.~Chen, Y.~Ju, Numerical experiments on reaction front
    propagation in n-heptane/air mixture with temperature gradient, Proceedings
    of the Combustion Institute 35~(3) (2015) 3045--3052.
    
    \bibitem{Yu2013}
    R.~Yu, X.-S. Bai, Direct numerical simulation of lean hydrogen/air
    auto-ignition in a constant volume enclosure, Combustion and Flame 160~(9)
    (2013) 1706--1716.
    
    \bibitem{Wang2013e}
    J.~Wang, Y.~Yang, Y.~Shi, Z.~Xiao, X.~He, S.~Chen, {Statistics and structures
        of pressure and density in compressible isotropic turbulence}, Journal of
    Turbulence 14~(6) (2013) 21--37.
    
    \bibitem{Donzis2013}
    D.~A. Donzis, S.~Jagannathan, {Fluctuations of thermodynamic variables in
        stationary compressible turbulence}, Journal of Fluid Mechanics 733 (2013)
    221--244.
    
    \bibitem{Jagannathan2016}
    S.~Jagannathan, D.~A. Donzis, Reynolds and mach number scaling in
    solenoidally-forced compressible turbulence using high-resolution direct
    numerical simulations, Journal of Fluid Mechanics 789 (2016) 669--707.
    
    \bibitem{Kuo1986}
    K.~K. Kuo, Principles of Combustion, John Wiley \& Sons, 1986.
    
    \bibitem{Fickett2000}
    W.~Fickett, W.~C. Davis, Detonation: theory and experiment, Courier
    Corporation, 2000.
    
    \bibitem{Kao2008}
    S.~Kao, J.~E. Shepherd, Numerical solution methods for control volume
    explosions and znd detonation structure, Tech. Rep. GALCIT Report FM2006.007,
    Aeronautics and Mechanical Engineering, California Institute of Technology,
    Pasadena, CA USA 91125 (2008).
    
    \bibitem{Sanchez2014}
    A.~L. S{\'a}nchez, F.~A. Williams, Recent advances in understanding of
    flammability characteristics of hydrogen, Progress in Energy and Combustion
    Science 41 (2014) 1--55.
    
    \bibitem{McBride1993}
    B.~J. McBride, S.~Gordon, M.~A. Reno, Coefficients for calculating
    thermodynamic and transport properties of individual species, Tech. rep.,
    NASA (1993).
    
    \bibitem{Hirschfelder1954}
    J.~O. Hirschfelder, C.~F. Curtiss, R.~B. Bird, M.~G. Mayer, Molecular theory of
    gases and liquids, Vol.~26, Wiley New York, 1954.
    
    \bibitem{Warnatz1982}
    J.~Warnatz, Influence of transport models and boundary conditions on flame
    structure, in: Numerical methods in laminar flame propagation, 1982, pp.
    87--111.
    
    \bibitem{Ern1994}
    A.~Ern, V.~Giovangigli, Multicomponent transport algorithms, Vol.~24, Springer
    Science \& Business Media, 1994.
    
    \bibitem{Kee1986}
    R.~J. Kee, G.~Dixon-Lewis, J.~Warnatz, M.~E. Coltrin, J.~A. Miller, A fortran
    computer code package for the evaluation of gas-phase multicomponent
    transport properties, Sandia National Laboratories Report SAND86-8246 13
    (1986) 80401--1887.
    
    \bibitem{Colella1990}
    P.~Colella, Multidimensional upwind methods for hyperbolic conservation-laws.,
    Journal of Computational Physics 87 (1990) 171--200.
    
    \bibitem{Gardiner2008}
    T.~A. Gardiner, J.~M. Stone, An unsplit {G}odunov method for ideal {MHD} via
    constrained transport in three dimensions., Journal of Computational Physics
    227 (2008) 4123--4141.
    
    \bibitem{Stone2008}
    J.~M. Stone, T.~A. Gardiner, P.~Teuben, J.~F. Hawley, J.~B. Simon, Athena: a
    new code for astrophysical {MHD}, The Astrophysical Journal Supplement Series
    178~(1) (2008) 137.
    
    \bibitem{Kee1996}
    R.~J. Kee, F.~M. Rupley, E.~Meeks, J.~A. Miller, {CHEMKIN-III: A FORTRAN
        chemical kinetics package for the analysis of gas-phase chemical and plasma
        kinetics}, {Sandia National Laboratories Report SAND96-8216}.
    
    \bibitem{Perini2012}
    F.~Perini, E.~Galligani, R.~D. Reitz, An analytical jacobian approach to sparse
    reaction kinetics for computationally efficient combustion modeling with
    large reaction mechanisms, Energy Fuels 26~(8) (2012) 4804--4822.
    
    \bibitem{Hamlington2017}
    P.~E. Hamlington, R.~Darragh, C.~A. Briner, C.~A.~Z. Towery, A.~Y. Poludnenko,
    Lagrangian analysis of high-speed turbulent premixed reacting flows:
    thermochemical trajectories in hydrogen-air flames, Combustion and Flame.
    
    \bibitem{Khokhlov2012}
    A.~Khokhlov, I.~Dom{\'\i}nguez, C.~Bacon, B.~Clifford, E.~Baron, P.~Hoeflich,
    K.~Krisciunas, N.~Suntzeff, L.~Wang, Three-dimensional simulations of
    thermonuclear detonation with alpha-network: Numerical method and preliminary
    results, in: Advances in Computational Astrophysics: Methods, Tools, and
    Outcome, Vol. 453, 2012, p. 107.
    
    \bibitem{Petersen2010}
    M.~R. Petersen, D.~Livescu, Forcing for statistically stationary compressible
    isotropic turbulence, Physics of Fluids 22~(11) (2010) 116101.
    
    \bibitem{Wang2017c}
    J.~Wang, T.~Gotoh, T.~Watanabe, Spectra and statistics in compressible
    isotropic turbulence, Physical Review Fluids 2~(1) (2017) 013403.
    
    \bibitem{Ristorcelli1997}
    J.~R. Ristorcelli, G.~A. Blaisdell, {Consistent Initial Conditions for the DNS
        of Compressible Turbulence}, Physics of Fluids 9~(1) (1997) 4--6.
    
    \bibitem{Samtaney2001}
    R.~Samtaney, D.~I. Pullin, B.~Kosovi\'{c}, {Direct numerical simulation of
        decaying compressible turbulence and shocklet statistics}, Physics of Fluids
    13~(5) (2001) 1415--1430.
    
    \bibitem{Lewiner2003}
    T.~Lewiner, H.~Lopes, A.~W. Vieira, G.~Tavares, Efficient implementation of
    marching cubes' cases with topological guarantees, Journal of Graphics Tools
    8~(2) (2003) 1--15.
    
    \bibitem{VanderWalt2014}
    S.~Van~der Walt, J.~L. Sch{\"o}nberger, J.~Nunez-Iglesias, F.~Boulogne, J.~D.
    Warner, N.~Yager, E.~Gouillart, T.~Yu, the scikit-image contributors,
    {scikit-image: image processing in Python}, PeerJ 2 (2014) e453.
    
    \bibitem{Akima1970}
    H.~Akima, A new method of interpolation and smooth curve fitting based on local
    procedures, Journal of the Association of Computing Machines 17~(4) (1970)
    589--602.
    
\end{thebibliography}

\end{document}